\documentclass{article}

\usepackage{arxiv}

\usepackage[utf8]{inputenc} 
\usepackage[T1]{fontenc}    
\usepackage[colorlinks,bookmarksopen,bookmarksnumbered,citecolor=cyan,urlcolor=cyan]{hyperref}            
\usepackage{url}            
\usepackage{booktabs}       
\usepackage{amsfonts}       
\usepackage{nicefrac}       
\usepackage{microtype}      
\usepackage{lipsum}		
\usepackage{graphicx}
\usepackage{natbib}
\usepackage{doi}
\usepackage{xurl}
\usepackage{xcolor}
\hypersetup{breaklinks=true}
\usepackage{mathrsfs}
\usepackage{amsmath}
\usepackage{amsfonts}
\usepackage{graphicx}
\usepackage{amsthm}
\usepackage{amssymb}
\usepackage{xtab,booktabs}
\usepackage{longtable}
\usepackage{caption}

\title{Heri-Graphs: A Workflow of Creating Datasets for Multi-modal Machine Learning on Graphs of Heritage Values and Attributes with Social Media}

\author{ \href{https://orcid.org/0000-0001-7637-3629}{\includegraphics[scale=0.06]{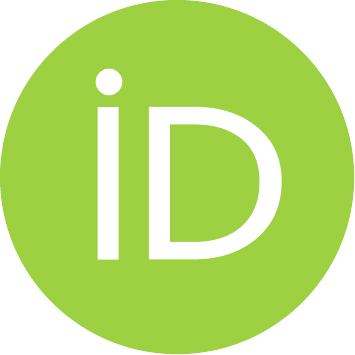}\hspace{1mm}Nan Bai}\\
	UNESCO Chair in Heritage and Values\\
	Delft University of Technology\\
	Delft, the Netherlands\\
	\texttt{n.bai@tudelft.nl} \\
	\And
	\href{https://orcid.org/0000-0002-3817-7931}{\includegraphics[scale=0.06]{orcid.pdf}\hspace{1mm}Pirouz Nourian} \\
	Genesis Lab | Chair of Design Informatics\\
	Delft University of Technology\\
	Delft, the Netherlands\\
	\texttt{p.nourian@tudelft.nl} \\
	\And
	\href{https://orcid.org/0000-0002-9062-3484}{\includegraphics[scale=0.06]{orcid.pdf}\hspace{1mm}Renqian Luo} \\
	Machine Learning Group\\
	Microsoft Research Asia\\
	Beijing, China\\
	\texttt{renqianluo@microsoft.com} \\
	\And
	\href{https://orcid.org/0000-0003-2571-9882}{\includegraphics[scale=0.06]{orcid.pdf}\hspace{1mm}Ana Pereira Roders} \\
	UNESCO Chair in Heritage and Values\\
	Delft University of Technology\\
	Delft, the Netherlands\\
	\texttt{a.r.pereira-roders@tudelft.nl} \\
}

\date{}

\renewcommand{\shorttitle}{HeriGraphs: A Workflow of Creating Datasets}

\hypersetup{
pdftitle={Heri-Graphs: A Workflow of Creating Datasets for Multi-modal Machine Learning on Graphs of Heritage Values and Attributes with Social Media},
pdfauthor={Nan Bai, Pirouz Nourian, Renqian Luo, Ana Pereira Roders},
pdfkeywords={World Heritage, Flickr, Multi-modal Dataset, Graph Construction, Machine and Deep Learning},
}

\begin{document}
\maketitle

\begin{abstract}
Values (why to conserve) and Attributes (what to conserve) are essential concepts of cultural heritage. Recent studies have been using social media to map values and attributes conveyed by public to cultural heritage.
However, it is rare to connect heterogeneous modalities of images, texts, geo-locations, timestamps, and social network structures to mine the semantic and structural characteristics therein. 
This study presents a methodological workflow for constructing such multi-modal datasets using posts and images on Flickr for graph-based machine learning (ML) tasks concerning heritage values and attributes. 
After data pre-processing using state-of-the-art ML models, the multi-modal information of visual contents and textual semantics are modelled as node features and labels, while their social relationships and spatiotemporal contexts are modelled as links in Multi-Graphs.
The workflow is tested in three cities containing UNESCO World Heritage properties - \emph{Amsterdam}, \emph{Suzhou}, and \emph{Venice}, which yielded datasets with high consistency for semi-supervised learning tasks.
The entire process is formally described with mathematical notations, ready to be applied in provisional tasks both as ML problems with technical relevance and as urban/heritage study questions with societal interests.
This study could also benefit the understanding and mapping of heritage values and attributes for future research in global cases, aiming at inclusive heritage management practices.\footnote{Data available at \href{https://github.com/zzbn12345/Heri_Graphs}{https://github.com/zzbn12345/Heri\_Graphs}}
\end{abstract}

\keywords{World Heritage \and Flickr \and Multi-modal Dataset \and Graph Construction \and Machine and Deep Learning}

\section{Introduction}\label{sec:Introduction}

In the context of UNESCO World Heritage (WH) Convention, ``\emph{values}" (why to conserve) and ``\emph{attributes}" (what to conserve) have been used extensively to detail the cultural significance of heritage \citep{UNESCO1972,UNESCO2008}.
Meanwhile, researchers have provided categories and taxonomies for heritage values and attributes, respectively \citep{PereiraRoders2007,TarrafaSilva2010,Veldpaus2015}.
Both concepts are essential for understanding the significance and meaning of cultural and natural heritage, and for making more comprehensive management plans \citep{Veldpaus2015}.
However, the heritage values and attributes are not only to define the significance of Outstanding Universal Value (OUV) in the particular context of World Heritage List (WHL), but all kinds of significance, ranging from listed to unlisted, natural to cultural, tangible to intangible, and from global to national, regional and local \citep{Rakic2008,TarrafaSilva2010,Bonci2018305,PereiraRoders2019,bai2021_2}.
Moreover, the 2011 UNESCO \emph{Recommendation on the Historic Urban Landscape} (HUL) stressed that heritage should also be recognized through the lens of local citizens, tourists and experts, calling for tools for civic engagement and knowledge documentation \citep{UNESCO2011,PereiraRoders2019,bai2021_2}.

Thereafter, in the past decade, analyses have been performed on User-Generated Content (UGC) from social media platforms to actively collect opinions of the [online] public, and to map heritage values and attributes conveyed by various stakeholders in urban environments \citep{Lu2015,Pickering2018}.
In Machine Learning (ML) literature, a \emph{modality} is defined as ``\emph{the way in which something happens or is experienced}'', which can include natural \emph{language}, \emph{visual} contents, \emph{vocal} signals, etc. \citep{Baltrusaitis2019}.
Most of studies mapping heritage values and attributes from UGC focused only on a few isolated modalities, such as textual topics of comments and/or tags \citep{Marine-Roig2015,Amato2016577,lee2021mining}, visual contents of depicted scenes \citep{zhang2019discovering,Giglio2019306}, social interactions \citep{Campillo2019,Liew2014,Williams201787}, and geographical distribution of the posts \citep{Giglio2019,Gabrielli2014}.

However, the heterogeneous multi-modal information from social media can enrich the understanding of posts, as textual and visual contents, temporal and geographical contexts, and underlined social network structures could show both complementary and contradictory messages \citep{Aggarwal2011,bai2021_2}.
A few studies have analysed different modalities to reveal the discussed topics and depicted scenes about cultural heritage \citep{Monteiro2014,Ginzarly2019}.
However, since they (mostly) adapted analogue approaches during analyses and the multi-modal information was not explicitly paired, linked, and analysed together, these studies could not yet be classified as \emph{Multi-modal Machine Learning} (MML), aiming to ``\emph{build models that can process and relate information from multiple modalities}'' \citep{Baltrusaitis2019} to enrich the conclusions that could not be easily achieved with isolated modalities.
On the other hand, \cite{Crandall2009} proposed a global dataset collected from Flickr with visual and textual features, as well as geographical locations.
Graphs were constructed with multi-modal information to map, cluster, and retrieve the most representative landmark images for major global cities.
\cite{Gomez2019530} trained multi-modal representation models of images, captions, and neighbourhoods with Instagram data within Barcelona, able to retrieve the most relevant photos and topics for each municipal district, being used to interpret the urban characteristics of different neighbourhoods.
More recently, the continuous research line demonstrated in \cite{kang2021transfer} and \cite{cho2022classifying} applied transfer learning \citep{pan2009survey} techniques to classify geo-tagged images into hierarchical scene categories and connected the depicted tourist activities to the urban environments that these cultural activities took place.
Although not all of them explicitly referred to heritage, these studies could provide useful information for scholars and practitioners to gather knowledge from the public about their perceived heritage values and attributes in urban settings, as suggested by HUL \citep{UNESCO2011, bai2021_2}.
Among the five main MML challenges summarized by \cite{Baltrusaitis2019}, representation (\emph{to present and summarize multi-modal data in a joint or coordinated space}) and fusion (\emph{to join information for prediction}) can be the most relevant for heritage and urban studies, as to 1) retrieving visual and/or textual information related to certain heritage values and attributes, 
and 2) aggregating individual posts in different geographic and administrative levels as the collective summarized knowledge of a place.

Furthermore, according to the \emph{First Law of Geography} \citep{tobler1970computer}, ``\emph{everything is related to everything else, but near things are more related than distant things}''.
This argument can also be assumed to be valid in other distance measures other than geographical ones where a random walk could be performed \citep{pearson1905problem}, such as in a topological space abstracted from spatial structure \citep{batty2013new, nourian2016configraphics,ren2019deep,zhang2020graph} or a social network constructed based on common interests \citep{wasserman1994social, lazer2009social,barabasi2013network, pentland2015social}.
In this light, it would be beneficial to construct graphs of UGC posts where Social Network Analysis (SNA) could be performed, showing the socio-economic and spatio-temporal context among them, reflecting the inter-related dependent nature of the posts \citep{cheng2014event}. 
Such a problem definition could help with both the classification and the aggregation tasks mentioned above, as has been demonstrated as effective and powerful by applications in the emerging field of Machine and Deep Learning on Graphs \citep{zhang2020deep, ma2021deep}.

The main contributions of this manuscript could be summarized as:
\begin{enumerate}
\item	Domain-specific multi-modal social network datasets about heritage values and attributes (or more precisely, the values and attributes conveyed by public to urban cultural heritage) are collected and structured with the User-Generated Content from the social media platform Flickr in three cities (Amsterdam, Suzhou, and Venice) containing UNESCO World Heritage properties, which could benefit the knowledge documentation and mapping for heritage and urban studies, aiming at a more inclusive heritage management process;
\item	State-of-the-art machine and deep learning models have been extensively applied and tested for generating multi-modal features and [pseudo-]labels with full mathematical formulations as its problem definition, providing a reproducible methodological workflow that could also be tested in other cases worldwide;
\item	Multi-graphs have been constructed to reflect the temporal, spatial, and social relationships among the data samples of collected User-Generated Content, ready to be further tested on several provisional tasks with both scientific relevance for Graph-based Multi-modal Machine Learning and Social Network research, and societal interests for Urban Studies, Urban Data Science, and Heritage Studies.
\end{enumerate}

\section{Materials and Methods}\label{sec:Material}

\subsection{Selection of Case Studies}\label{sec:case study}
Without loss of generality, this research selected three cities in Europe and China that are related to UNESCO WH and HUL as case studies: Amsterdam (AMS), the Netherlands; Suzhou (SUZ), China; and Venice (VEN), Italy.
All three cities either are themselves entirely or partially inscribed in the WHL, such as \emph{Venice and its Lagoon\footnote{\href{https://whc.unesco.org/en/list/394}{https://whc.unesco.org/en/list/394}}} and \emph{Seventeenth-Century Canal Ring Area of Amsterdam inside the Singelgracht\footnote{\href{https://whc.unesco.org/en/list/1349/}{https://whc.unesco.org/en/list/1349/}}}, or contain WHL in multiple spots of the city, such as \emph{Classical Gardens of Suzhou\footnote{\href{http://whc.unesco.org/en/list/813}{http://whc.unesco.org/en/list/813}}}, showcasing different spatial typologies of cultural heritage in relation to its urban context \citep{PereiraRoders2010, Valese2020}.

As shown in Table~\ref{T_Case_studies}, the three cases have very different scales, yet they all strongly demonstrated the relationship of urban fabric and water system.
Interestingly, Amsterdam and Suzhou have been respectively referred to as ``\emph{the Venice of the North/East}"
by media and public.
Moreover, 
the concept of OUV introduced in Section~\ref{sec:Introduction} reveals the core heritage values of WH properties.
The OUV of a property would be justified with ten selection criteria, where criteria (i)-(vi) reflect various cultural values, and criteria (vii)-(x) natural ones \citep{jokilehto2007ouv,Jokilehto2008,UNESCO2008,Bai2021}, as explained in Appendix Table~\ref{T_OUV_definition}.
The three selected cases include a broad variety of all cultural heritage OUV selection criteria, implying the representativeness of the datasets constructed in this study.

\begin{table*}[ht]
\scriptsize\centering
\caption{\footnotesize The case studies and their World Heritage status.\label{T_Case_studies}}
\sf\begin{tabular}{p{40pt}p{50pt}p{140pt}lrr}
\toprule
City & Geo-location & WHL Name & OUV Criteria & Area of Property & Date of Inscription\\
\midrule
Amsterdam (AMS) & 52.365000N 4.887778E & \emph{Seventeenth-Century Canal Ring Area of Amsterdam inside the Singelgracht} & (i)(ii)(iv) & 198.2 ha & 2010\\
Suzhou (SUZ) & 31.302300N 120.631300E & \emph{Classical Gardens of Suzhou} & (i)(ii)(iii)(iv)(v) & 11.9 ha & 2000\\
Venice (VEN) & 45.438759N 12.327145E & \emph{Venice and its Lagoon} & (i)(ii)(iii)(iv)(v)(vi) & 70,176.2 ha & 1987\\
\bottomrule
\end{tabular}\\
\end{table*}

\subsection{Data Collection and Pre-processing}\label{sec:data collection}
Numerous studies have collected, annotated, and distributed open-source datasets from the social media platform Flickr due to its convenient Application Programming Interface (API), including \emph{MirFlickr-1M} \citep{Huiskes2008}, \emph{NUS-WIDE} \citep{Chua2009}, \emph{Flickr} \citep{tang2009relational}, \emph{ImageNet} \citep{deng2009imagenet, krizhevsky2012imagenet}, Microsoft Common Object in COntext (\emph{MS COCO}) \citep{lin2014microsoft}, \emph{Flickr30k} \citep{plummer2015flickr30k}, \emph{SinoGrids} \citep{zhou2016sinogrids}, and \emph{GRAPH Saint} \citep{zeng2019graphsaint}, etc.
These datasets containing one or more of the visual, semantic, social, and/or geographical information of UGC are widely used, tested, but also sometimes challenged by different ML communities including \emph{Computer Vision}, \emph{Multi-modal Machine Learning}, and \emph{Machine Learning on Graphs}.
However, they are more or less suitable for bench-marking general ML tasks and testing computational algorithms, which are not necessarily tailor-made for heritage and urban studies.
The motivation of data collection in this research is to provide datasets that could be both directly applicable for ML communities as test-bed, and theoretically informative for heritage and urban scholars to draw conclusions on for planning decision-making.

FlickrAPI python library\footnote{\href{https://stuvel.eu/software/flickrapi/}{https://stuvel.eu/software/flickrapi/}} was used to access the API method provided by Flickr\footnote{\href{https://www.flickr.com/services/api/}{https://www.flickr.com/services/api/}}, using the Geo-locations in Table~\ref{T_Case_studies} as the centroids to search a maximum of 5000 IDs of geo-tagged images in a fixed radius covering the major urban area, to make the datasets from the three cities comparable and compatible.
To test the scalability of the methodological workflow, another larger dataset without ID number limit has also been collected in Venice (VEN-XL).
Only images with a \texttt{\small candownload} flag indicated by the owner were further queried, respecting the privacy and copyrights of Flickr users.
The following information of each post was collected: owner's ID; owner's registered location on Flickr; the title, description, and tags provided by user; geo-tag of the image; timestamp marking when the image was taken, and URLs to download the \texttt{\small Large Square} (150$\times$150 px) and \texttt{\small Small 320} (320$\times$240 px) versions of the original image.
Furthermore, the public friend and subscription lists of all the retrieved owners have been queried, while all personal information were only considered as a [semi-] anonymous ID as respect to the privacy policy. 

The retrieved textual fields of \texttt{\small description}, \texttt{\small title}, and \texttt{\small tags} could all provide useful information, yet not all posts have these fields, and not all posts are necessarily written to express thoughts and share knowledge about the place (considered as \emph{valid} in the context of this study).
The textual fields of the posts were cleaned, translated, and merged into a \texttt{\small Revised Text} field as the raw English textual data, after recording the detected original language of posts on sentence level using Google Translator API from the Deep Translator python library\footnote{\href{https://deep-translator.readthedocs.io/en/latest/}{https://deep-translator.readthedocs.io/en/latest/}}.
Moreover, many posts shared by the same user were uploaded at once, thus having the same duplicated textual fields for all of them.
To handle such redundancy, a separate dataset of all the unique processed textual data on sentence level was saved for each city, while the original post of each sentence was marked and could easily be traced back.

Detailed description of the data collection and pre-processing methods could be found in Appendix~\ref{sec:App_data_collection}.
Table~\ref{T_data_collection} shows the number of data samples (posts) and owners (users) for the three case study cities at each stage.

\begin{table}[ht]
\scriptsize\centering
\caption{\footnotesize The number of data samples collected at each stage.}\label{T_data_collection}
\sf
\begin{tabular}{lrrrr}
\toprule
City & AMS & SUZ & VEN & VEN-XL\\
\midrule
IDs Collected & 5000 & 4229 & 5000 & 116,675\\
Is Downloadable & 3727 & 3137 & 2952 & 80,964\\
\textbf{Downloaded Posts}* & \textbf{3727} & \textbf{3137} & \textbf{2951} & \textbf{80,963}\\\midrule
Has Textual Data** & 3404 & 2692 & 2801 & 77,644\\
Has Unique Texts*** & 3130 & 1963 & 1952 & 59,396\\
Unique Sentences & 2247 & 361 & 3249 & 61,253\\
Original Posts*** & 2904 & 754 & 1761 & 49,823\\
\midrule
Posting Owners & 195 & 95 & 330 & 6077\\
\bottomrule
\multicolumn{5}{l}{\scriptsize *Regarded as the sample size of the dataset.}\\
\multicolumn{5}{l}{\scriptsize **At least one of \texttt{\scriptsize Description}, \texttt{\scriptsize Title} and \texttt{\scriptsize Tag} fields is not empty.}\\
\multicolumn{5}{l}{\scriptsize ***Different because of posts without any \emph{valid} sentences.}
\end{tabular}
\end{table}

\begin{figure*}[ht]
\centering
\includegraphics[width=\linewidth]{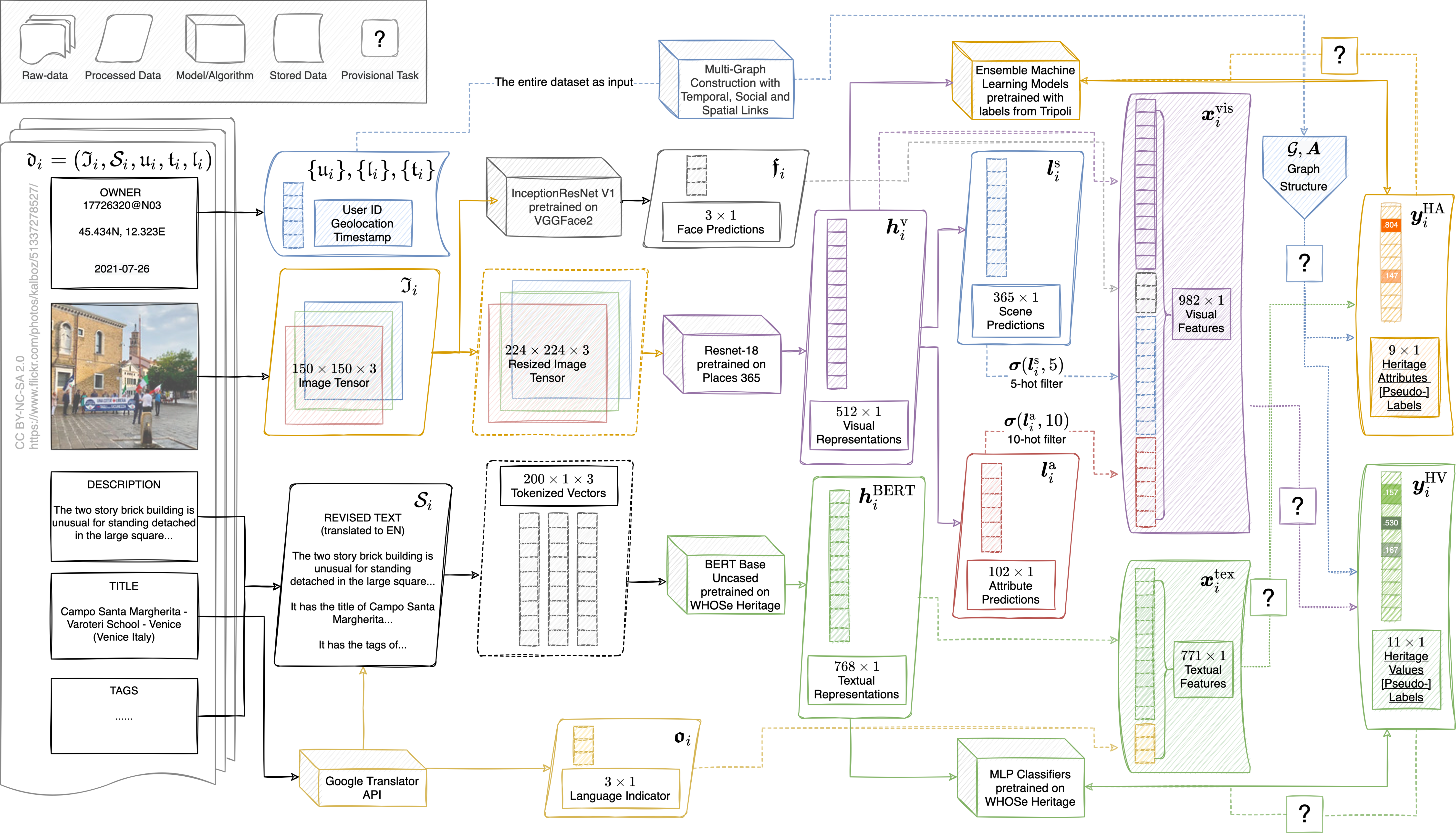}
\caption{\footnotesize The workflow of multi-modal feature generation process of one sample post in Venice, while graph construction requires all data points of the dataset. The \href{https://www.flickr.com/photos/kalboz/51337278527/
}{original post} owned by user \texttt{\small 17726320@N03} is under \href{https://creativecommons.org/licenses/by-nc-sa/2.0/}{CC BY-NC-SA 2.0} license. The question marks in the right part indicate some provisional tasks for this dataset, which will be discussed in Section~\ref{sec:urban data science} and Table~\ref{T_provisional_tasks}.}\label{fig:DataFlow}
\end{figure*}

To formally describe the data, define the problem, and propose a generalizable workflow, mathematical notations are used in the rest of this manuscript.
Since the same process is valid for all three cities (and probably also for other unselected cases worldwide) and has been repeated exactly three times, no distinctions would be made among the cities, except for the cardinality of sets reflecting sample sizes.
Let $i$ be the index of a generic sample of the dataset for one city, then the raw data of it could be denoted as a tuple $\boldsymbol{\mathfrak{d}}_i = (\boldsymbol{\mathfrak{I}}_i, \mathcal{S}_i, \mathfrak{u}_i, \mathfrak{t}_i, \mathfrak{l}_i), \boldsymbol{\mathfrak{d}}_i \in \mathfrak{D}= \{\boldsymbol{\mathfrak{d}}_1, \boldsymbol{\mathfrak{d}}_2,..., \boldsymbol{\mathfrak{d}}_K\}$, where $K$ is the sample size of the dataset in a city (as shown in Table~\ref{T_data_collection}), $\boldsymbol{\mathfrak{I}}_i$ is a three-dimensional tensor of the size of the image with three RGB channels, $\mathcal{S}_i=\{\mathcal{s}^{(1)}_i, \mathcal{s}^{(2)}_i, ..., \mathcal{s}^{(|\mathcal{S}_i|)}_i\} \text{ or } \mathcal{S}_i=\varnothing $ is a set of revised English sentences that can also be an empty set for samples without any valid textual data, $\mathfrak{u}_i \in \mathcal{U}$ is a user ID that is one instance from the user set $\mathcal{U}=\{\mu_1, \mu_2, ..., \mu_{|\mathcal{U}|}\}$, $\mathfrak{t}_i \in \mathcal{T}$ is a timestamp that is one instance from the ordered set of all the unique timestamps $\mathcal{T}=\{\tau_1, \tau_2, ..., \tau_{|\mathcal{T}|}\}$ from the dataset at the level of weeks, and $\mathfrak{l}_i = (\mathfrak{x}_i,\mathfrak{y}_i)$ is a geographical coordinate of latitude ($\mathfrak{y}_i$) and longitude ($\mathfrak{x}_i$) marking the geo-location of the post.
A complete nomenclature of all the notations used in this paper can be found in the Appendix Tables~\ref{T_nomenclature} and \ref{T_func}.
Figure~\ref{fig:DataFlow} demonstrates the workflow of one sample post in Venice, which will be explained in the following sections.

\subsection{Multi-modal Features Generation}\label{sec:feature generation}



\subsubsection{Visual Features} \label{sec:vis feature}
\emph{Places365} is a dataset that contains 1.8 million images from 365 scene categories, which includes a relatively comprehensive collection of indoor and outdoor places \citep{zhou2014learning,zhou2017places}.
The categories can be informative for urban and heritage studies to identify depicted scenes of images and to further infer heritage attributes \citep{Veldpaus2015,Ginzarly2019}.
A few Convolutional Neural Network (CNN) models have been pretrained by \cite{zhou2017places} using state-of-the-art backbones to predict the depicted scenes in images, reaching a top-1 accuracy of around 55\% and top-5 accuracy of around 85\%.
Furthermore, the same set of pretrained models have been used to predict 102 discriminative scene attributes based on \emph{SUN Attribute} dataset \citep{patterson2012sun, patterson2014sun}, reaching top-1 accuracy of around 92\% \citep{zhou2017places}.
These scene attributes are conceptually different from heritage attributes, as the former are mostly adjectives and present participles describing the scene and activities taking place, therefore both heritage values and attributes could be effectively inferred therefrom.

This study used the openly-released ResNet-18 model \citep{he2016deep} pretrained on \emph{Places365} with PyTorch\footnote{\href{https://github.com/CSAILVision/places365}{https://github.com/CSAILVision/places365}}.
This model was adjusted to effectively yield three output vectors: 1) the last softmax layer of the model $\boldsymbol{l}^\text{s}_{365\times1}$ as logits over all scene categories; 2) the last hidden layer $\boldsymbol{h}^\text{v}_{512\times1}$ of the model; 3) a vector $\boldsymbol{l}^\text{a}_{102\times1}$ as logits over all scene attributes.
Such a process for any image input $\boldsymbol{\mathfrak{I}}_i$ could be described as:
\begin{equation}
\label{eq:CNN_sample}
    \boldsymbol{l}^\text{s}_i, \boldsymbol{l}^\text{a}_i, \boldsymbol{h}^\text{v}_i=\boldsymbol{f}_\text{ResNet-18}(\boldsymbol{\mathfrak{I}}_i|\boldsymbol{\Theta}_\text{ResNet-18}),
\end{equation}
or preferably in a vectorized format:
\begin{equation}
\label{eq:CNN_v}
    \boldsymbol{L}^\text{s}, \boldsymbol{L}^\text{a}, \boldsymbol{H}^\text{v}=\boldsymbol{f}_\text{ResNet-18}([\boldsymbol{\mathfrak{I}}_1, \boldsymbol{\mathfrak{I}}_2,..., \boldsymbol{\mathfrak{I}}_K]|\boldsymbol{\Theta}_\text{ResNet-18}), 
\end{equation}
where
\begin{equation}
\label{eq:CNN_vs}
    \boldsymbol{L}^\text{s}:=[\boldsymbol{l}^\text{s}_i]_{365\times K}, 
    \boldsymbol{L}^\text{a}:=[\boldsymbol{l}^\text{a}_i]_{102\times K},
    \boldsymbol{H}^\text{v}:=[\boldsymbol{h}^\text{v}_i]_{512\times K}.
\end{equation}
Considering that the models have reasonable performance in top-$n$ accuracy, to keep the visual features explainable, a $n$-hot soft activation filter $\boldsymbol{\sigma}^{(n)}$ is performed on both logit outputs, to keep the top-$n$ prediction entries active, while smoothing all the others based on the confidence of top-$n$ predictions ($n=5$ for scene categories $\boldsymbol{L}^\text{s}$ and $n=10$ for scene attributes $\boldsymbol{L}^\text{a}$).
Let $\text{max}(\boldsymbol{l}, n)$ denote the $n_\text{th}$ maximum element of a $d$-dimensional logit vector $\boldsymbol{l}$ (the sum of all entries of $\boldsymbol{l}$ equals 1), then the activation filter $\boldsymbol{\sigma}^{(n)}$ could be described as:
\begin{align}
\label{eq:Active_filter}
    \boldsymbol{\sigma}^{(n)}(\boldsymbol{l}_{d\times1})=\boldsymbol{l}\odot\boldsymbol{m} + \frac{1-\boldsymbol{l}^\mathsf{T}\boldsymbol{m}}{d-n} (\boldsymbol{1}_{d\times1} -\boldsymbol{m}),\\
    \boldsymbol{m}:=[m_\iota]_{d\times 1}, m_\iota=
    \begin{cases}
      1 & \text{if } l_\iota \geq \text{max}(\boldsymbol{l}, n)\\
      0 & \text{otherwise}
    \end{cases},
\end{align}
where $\boldsymbol{m}$ is a mask vector indicating the positions of top-$n$ entries, and $\boldsymbol{l}^T\boldsymbol{m}$ is effectively the total confidence of the model for top-$n$ predictions.
Note that this function could also take a matrix as input and process it as several column vectors to be concatenated back.

Furthermore, as \emph{Places365} dataset is tailor-made for scene detection tasks rather than recognizing faces \citep{zhou2017places}, the models pretrained on it may get confused when a new image is mainly composed of faces as ``typical tourism pictures'' and selfies, which is not uncommon in the case studies as popular tourism destinations.
As the ultimate aim of constructing such datasets is not to precisely predict the scene each image depict, but to help infer heritage values and attributes, it would be unfair to simply exclude those images with significant proportion of faces on them.
Rather, the existence of human on the images showing their activities would be a strong cue of intangible dimension of heritage properties.
Under such consideration, an Inception ResNet-V1 model\footnote{\href{https://github.com/timesler/facenet-pytorch}{https://github.com/timesler/facenet-pytorch}}  pretrained on the \emph{VGGFace2} Dataset \citep{schroff2015facenet,cao2018vggface2} has been used to generate features about depicted faces in the images.
A three dimensional vector $\boldsymbol{\mathfrak{f}}_i$ was obtained for any image input $\boldsymbol{\mathfrak{I}}_i$, where the non-negative first entry $\mathfrak{f}_{1,i}\in\mathbb{N}$ counts the number of faces detected in the image, the second entry $\mathfrak{f}_{2,i}\in [0,1]$ records the confidence of the model for face detection, and the third entry $\mathfrak{f}_{3,i}\in [0,1]$ calculates the proportion of area of all the bounding boxes of detected faces to the total area of the image.
Similarly, the vectorized format could be written as $\boldsymbol{F}:=[\boldsymbol{\mathfrak{f}}_i]_{3\times K}$ over the entire dataset.

Finally, all the obtained visual features were concatenated vertically to generate the final visual feature $\boldsymbol{X}^\text{vis}_{982\times K}$:
\begin{equation}
\label{eq:Vis Feat}
    \boldsymbol{X}^\text{vis}_{982\times K} = {\left[{\boldsymbol{H}^\text{v}}^\mathsf{T}, {\boldsymbol{F}}^\mathsf{T}, {\boldsymbol{\sigma}^{(5)}(\boldsymbol{L}^\text{s})}^\mathsf{T}, {\boldsymbol{\sigma}^{(10)}(\boldsymbol{L}^\text{a})}^\mathsf{T}\right]}^\mathsf{T},
\end{equation}
where $[\cdot,\cdot]$ denotes the horizontal concatenation of matrices.

This final matrix is to be used in future MML tasks as the vectorized descriptor of the uni-modal visual contents of the posts, with both more abstract hidden features, and more specific information about predicted categories, which is a common practice in MML literature \citep{Baltrusaitis2019}.
All models are tested on both the $150\times 150$ and $320\times 240$ px images to compare the consistency of generated features.
The workflow of generating visual features is illustrated in the top part of Figure~\ref{fig:DataFlow}.

\subsubsection{Textual Features} \label{sec: tex feature}
In the last decade, attention- and Transformer-based models have taken over the field of Natural Language Processing (NLP), increasing the performance of models in both general machine learning tasks, and domain-specific transfer-learning scenarios \citep{vaswani2017attention}.
As an early version, the pretrained Bidirectional Encoder Representations from Transformers (BERT) \citep{devlin2019bert} is still regarded as a powerful base model to be fine-tuned on specific downstream datasets and to perform various NLP tasks.
Specifically, the output on the \texttt{\small [CLS]} token of BERT models is regarded as an effective representation of the entire input sentence, being used extensively for classification tasks \citep{Clark2019,Sun2019}.
In heritage studies domain, \cite{bai-etal-2021-whose-heritage} fine-tuned BERT on the dataset \emph{WHOSe Heritage} they constructed from UNESCO World Heritage inscription document, followed by a Multi-Layer Perceptron (MLP) classifier to predict the OUV selection criteria a sentence is concerned with, showing top-1 accuracy of around 71\% and top-3 accuracy of around 94\%.

This study used the openly-released BERT model fine-tuned on \emph{WHOSe Heritage} with PyTorch\footnote{\href{https://github.com/zzbn12345/WHOSe_Heritage}{https://github.com/zzbn12345/WHOSe\_Heritage}}.
The BERT model took both the entire sentence sets $\mathcal{S}_i$ and individual sentences of the sets $\{\mathcal{s}^{(1)}_i, \mathcal{s}^{(2)}_i, ..., \mathcal{s}^{(|\mathcal{S}_i|)}_i\}$ as paragraph-level and sentence-level inputs, respectively, for the comparison of its consistency on their predicted outputs on this new dataset.
Furthermore, taking the entire sentence sets $\mathcal{S}_i$ as input, the 768-dimensional output vector $\boldsymbol{h}^\text{BERT}_{768\times1}$ of the \texttt{\small [CLS]} token was retrieved on samples that have valid textual data:
\begin{equation}
\label{eq:BERT_sample}
    \boldsymbol{h}^\text{BERT}_i=
      \boldsymbol{f}_\text{BERT}(\mathcal{S}_i|\boldsymbol{\Theta}_\text{BERT}), \text{ where } \boldsymbol{f}_\text{BERT}(\varnothing|\boldsymbol{\Theta}_\text{BERT})=\boldsymbol{0}_{768\times 1}
\end{equation}
or preferably in a vectorized format:
\begin{align}
\label{eq:BERT_v}
    \boldsymbol{H}^\text{B}=\boldsymbol{f}_\text{BERT}([\mathcal{S}_1, \mathcal{S}_2,..., \mathcal{S}_K]|\boldsymbol{\Theta}_\text{BERT}),
\text{where }
    \boldsymbol{H}^\text{B}:=[\boldsymbol{h}^\text{BERT}_i]_{768\times K}.
\end{align}

Moreover, the original language of each sentence may provide additional information to the understanding of the verbal context of posts, which can also be informative to effectively identify and compare locals and tourists.
A three dimensional vector $\boldsymbol{\mathfrak{o}}_i \in \{0,1\}^3$ was obtained with Google Translator API.
The three entries respectively marked whether there were sentences in English, local languages (Dutch, Chinese, or Italian, respectively), and other languages in the set $\mathcal{S}_i$.
The elements of vector $\boldsymbol{\mathfrak{o}}_i$ or the matrix form $\boldsymbol{{O}}:=[\boldsymbol{\mathfrak{o}}_i]_{3\times K}$ could be in a range from all zeros (when there were no textual data at all) to all ones (when the post was composed of different languages in separate sentences).

Similar to visual features, final textual features $\boldsymbol{X}^\text{tex}_{771\times K}$ could be obtained by concatenation:
\begin{equation}
\label{eq:Tex Feat}
    \boldsymbol{X}^\text{tex}_{771\times K} = {\left[{\boldsymbol{H}^\text{B}}^\mathsf{T}, {\boldsymbol{O}}^\mathsf{T}\right]}^\mathsf{T}.
\end{equation}
The workflow of generating textual features is illustrated in the bottom part of Figure~\ref{fig:DataFlow}.

\subsubsection{Contextual Features} \label{sec:cntx feature}
As mentioned in Section~\ref{sec:data collection}, the user ID $\mathfrak{u}_i$ and timestamp $\mathfrak{t}_i$ of a post are both an instance from their respective set $\mathcal{U}$ and $\mathcal{T}$, since multiple posts could be posted by the same user, and multiple images could be taken during the same week.
To help formulate and generalize the problem under the practice of relational database \citep{reiter1989towards}, both information could be transformed as one-hot embeddings $\boldsymbol{U}:=[u_{j,i}]_{|\mathcal{U}| \times K}\in \{0,1\}^{|\mathcal{U}| \times K}$ and $\boldsymbol{T}:=[t_{k,i}]_{|\mathcal{T}| \times K}\in \{0,1\}^{|\mathcal{T}| \times K}$, such that:
\begin{align}
\label{eq:One-hot U}
    u_{j,i}=
    \begin{cases}
      1 & \text{if } \mathfrak{u}_i=\mu_j \in \mathcal{U}\\
      0 & \text{otherwise}
    \end{cases},\\
    \text{and }
\label{eq:One-hot T}
    t_{k,i}=
    \begin{cases}
      1 & \text{if } \mathfrak{t}_i=\tau_k \in \mathcal{T}\\
      0 & \text{otherwise}
    \end{cases}.
\end{align}

Furthermore, Section~\ref{sec:data collection} also mentioned the collection of the public contacts and groups of all the users $\mu_j$ from the set $\mathcal{U}$.
To keep the problem simple, only direct contact pairs were considered to model the back-end social structure of the users, effectively filtering out the other contacts a user $\mu_j$ has that were not in the set of interest $\mathcal{U}$, resulting an adjacency matrix among the users $\boldsymbol{A}^\mathcal{U}:=[a^\mathcal{U}_{j, j'}]_{|\mathcal{U}|\times |\mathcal{U}|} \in\{0,1\}^{|\mathcal{U}| \times |\mathcal{U}|}, j, j'\in[1,|\mathcal{U}|]$ marking their direct friendship:
\begin{equation}
    \label{eq:friendship}
    a^\mathcal{U}_{j,j'}=
    \begin{cases}
      1 & \text{if } \mu_{j} \text{ and } \mu_{j'} \text{ are contacts}\text{ or } j=j'\\
      0 & \text{otherwise}
    \end{cases}.\\
\end{equation}
Let $\mathcal{I}(\mu_j)$ denote the set of public groups a user $\mu_j$ follows (can be an empty set if $\mu_j$ follows no group), and let $\text{IoU}(\mathcal{A},\mathcal{B})$ denote the Jaccard Index (size of Intersection over size of Union) of two generic sets $\mathcal{A},\mathcal{B}$:
\begin{equation}
    \label{eq:IoU}
    \text{IoU}(\mathcal{A},\mathcal{B})=\frac{|\mathcal{A}\cap\mathcal{B}|}{|\mathcal{A}\cup\mathcal{B}|},
\end{equation}
then another weighted adjacency matrix among the users $\boldsymbol{A}^\mathcal{U'}:=[a^\mathcal{U'}_{j, j'}]_{|\mathcal{U}|\times |\mathcal{U}|} \in[0,1]^{|\mathcal{U}| \times |\mathcal{U}|}, j, j'\in[1,|\mathcal{U}|]$ could be constructed marking the mutual interests among the users in terms of group subscription on Flickr:
\begin{equation}
    \label{eq:interest}
    a^\mathcal{U'}_{j,j'}=
    \begin{cases}
      0 & \text{if } \mathcal{I}(\mu_{j})\cup \mathcal{I}(\mu_{j'})=\varnothing\\
      \text{IoU}(\mathcal{I}(\mu_{j}), \mathcal{I}(\mu_{j'})) & \text{otherwise}
    \end{cases}.\\
\end{equation}

To further simplify the problem, although the geo-location $\mathfrak{l}_i = (\mathfrak{x}_i,\mathfrak{y}_i)$ of each post was typically distributed in a continuous range in the 2D geographical space, it would be beneficial to further aggregate and discretize the distribution in a topological abstraction of spatial network \citep{batty2013new,nourian2016configraphics, nourian2016spectral}, which has also been proven to be effective in urban spatial analysis, including but not limited to Space Syntax \citep{hillier1989social,penn2003space,Ratti2004,blanchard2008mathematical}.
The OSMnx python library\footnote{\href{https://osmnx.readthedocs.io/en/stable/}{https://osmnx.readthedocs.io/en/stable/}} was used to inquire the simplified spatial network data on \emph{OpenStreetMap} including all means of transportation \citep{boeing2017osmnx} in each city with the same centroid location and radius described in Section~\ref{sec:data collection}.
This operation effectively saved a spatial network as an undirected weighted graph $G_0=(V_0,E_0,\boldsymbol{w}_0)$, where $V_0=\{\upsilon_1, \upsilon_2, ..., \upsilon_{|V_0|}\}$ is the set of all street intersection nodes, $E_0\subseteq V_0\times V_0$ is the set of all links possibly connecting two spatial nodes (by different sorts of transportation such as walking, biking, and driving), and $\boldsymbol{w}_0 \in \mathbb{R}_+^{|E_0|}$ is a vector with the same dimension as the cardinality of the edge set, marking the average travel time needed between node pairs (dissimilarity weights).
The \texttt{\small distance.nearest\_nodes} method of OSMnx library was used to retrieve the nearest spatial nodes to any post location $\mathfrak{l}_i = (\mathfrak{x}_i,\mathfrak{y}_i)$.
By only keeping the spatial nodes that have at least one data sample posted nearby, and restricting the link weights between nodes so that the travel time on any link is no more than 20 minutes, which ensures a comfortable temporal distance forming neighbourhoods and communities \citep{howley2009sustainability}, a subgraph $G=(V,E,\boldsymbol{w})$ of $G_0$ could be constructed, so that $V \subseteq V_0, E \subseteq E_0$, and $\boldsymbol{w} \in [0,20.0]^{|E|}$.
As the result, another one-hot embedding matrix $\boldsymbol{S}:=[s_{l,i}]_{|V| \times K}\in \{0,1\}^{|V| \times K}$ could be obtained:
\begin{equation}
\label{eq:One-hot Geo}
    s_{l,i}=
    \begin{cases}
      1 & \text{if the closest node to point } \mathfrak{l}_i \text{ is }\upsilon_l \in V\\
      0 & \text{otherwise}
    \end{cases}.
\end{equation}

The contextual features constructed as matrices/graphs would be further used in Section~\ref{sec:graph construction} to link the posts together.

\subsection{Pseudo-Label Generation}\label{sec:label_generation}
\subsubsection{Heritage Values as OUV Selection Criteria} \label{sec:value_label}
Various categories on heritage values (HV) have been provided by scholars \citep{PereiraRoders2007,jokilehto2007ouv,Jokilehto2008,TarrafaSilva2010}.
To keep the initial step simple, this study arbitrarily applied the value definition in UNESCO WHL with regard to ten OUV selection criteria, as listed in Appendix Table~\ref{T_OUV_definition} with an additional class \texttt{\small Others} representing scenarios where no OUV selection criteria suit the scope of a sentence (resulting a 11-class category).
A group of ML models have been trained and fine-tuned to make such predictions by \cite{bai-etal-2021-whose-heritage} as introduced in Section~\ref{sec: tex feature}.
Except for BERT already used to generate textual features as mentioned above, a Universal Language Model Fine-tuning (UMLFiT) \citep{ulmfit} has also been trained and fine-tuned, reaching a similar performance in accuracy.
Furthermore, it has been found that the average confidence by both BERT and ULMFiT models on the prediction task showed significant correlation with expert evaluation, even on social media data \citep{bai-etal-2021-whose-heritage}.
This suggests that it may be possible to use the both trained model to generate labels about heritage values in a semi-supervised active learning setting \citep{prince2004does,zhu2009introduction}, as this is a task too knowledge-demanding for crowd-workers, yet too time-consuming for experts \citep{pustejovsky2012natural}.

The pseudo-label generation step could be formulated as:
\begin{align}
\label{eq:value_label}
    \boldsymbol{y}^\text{BERT}_{i}=
    \begin{cases}
      \boldsymbol{g}_\text{BERT}(\mathcal{S}_i|\boldsymbol{\Theta}_\text{BERT})& \text{ if } \mathcal{S}_i \neq \varnothing\\
      \boldsymbol{0}_{11\times1}& \text{ otherwise}
    \end{cases},\\
    \boldsymbol{y}^\text{ULMFiT}_{i}=
    \begin{cases}
      \boldsymbol{g}_\text{ULMFiT}(\mathcal{S}_i|\boldsymbol{\Theta}_\text{ULMFiT})& \text{ if } \mathcal{S}_i \neq \varnothing\\
      \boldsymbol{0}_{11\times1}& \text{ otherwise}
    \end{cases},\\
    \boldsymbol{Y}^\text{HV} := [\boldsymbol{y}^\text{HV}_i]_{11\times K}, \boldsymbol{y}^\text{HV}_{i}= \frac{\boldsymbol{y}^\text{BERT}_{i}+\boldsymbol{y}^\text{ULMFiT}_{i}}{2}.
\end{align}
where $\boldsymbol{g}_*$ is an end-to-end function including both pre-trained models and MLP classifiers; and $\boldsymbol{y}^*_i$ is an 11-dimensional logit vector as soft-label predictions. 
Let $\text{argmx}(\boldsymbol{l},n)$ denote the function returning the index set of the largest $n$ elements of a vector $\boldsymbol{l}$, together with the previously defined $\text{max}(\boldsymbol{l},n)$,
the confidence and [dis-]agreement of models for top-$n$ predictions could be computed as:
\begin{align}
\label{eq:value_conf}
    \boldsymbol{K}^\text{HV} := [\boldsymbol{\kappa}^\text{HV}_i]_{2\times K},
    \boldsymbol{\kappa}^\text{HV}_i := [\kappa^{\text{HV}(0)}_i,\kappa^{\text{HV}(1)}_i]^\mathsf{T},\\
    \kappa^{\text{HV}(0)}_i= \sum_{n_0=1}^{n}\frac{\text{max}(\boldsymbol{y}^\text{BERT}_{i},n_0)+\text{max}(\boldsymbol{y}^\text{ULMFiT}_{i},n_0)}{2},\\
    \kappa^{\text{HV}(1)}_i= \text{IoU}(\text{argmx}(\boldsymbol{y}^\text{BERT}_{i},n), \text{argmx}(\boldsymbol{y}^\text{ULMFiT}_{i},n)).
\end{align}
This confidence indicator matrix $\boldsymbol{K}^\text{HV}$ could be presumably regarded as a filter for the labels on heritage values $\boldsymbol{Y}^\text{HV}$, to only keep the samples with high inter-annotator (model) agreement \citep{Nowak2010} as the ``ground-truth" [pseudo-] labels, while treating the others as unlabeled \citep{lee2013pseudo,sohn2020fixmatch}.

\subsubsection{Heritage Attributes as Depicted Scenery}\label{sec:attr_label}
Heritage attributes (HA) also have multiple categorization systems
\citep{veldpaus2014learning,Veldpaus2015,gustcoven2016attributes,Ginzarly2019,UNESCO2020}, and are arguably more vaguely defined than HV.
For simplicity, this study arbitrarily combined the attribute definitions of \cite{Veldpaus2015} and \cite{Ginzarly2019} and kept a 9-class category of tangible and/or intangible attributes that were visible from an image.
More precisely speaking, this category should be framed as ``depicted scenery" of an image \citep{Ginzarly2019} that heritage attributes could possibly be induced from.
The depicted scenes themselves are not yet valid heritage attributes.
This semantic/philosophical discussion, however, is out of the scope of this paper.
The definitions of the nine categories are listed in Appendix Table~\ref{T_attr_definition}.

An image dataset collected in Tripoli, Lebanon and classified with expert-based annotations presented by \cite{Ginzarly2019} was used to train state-of-the-art ML models to replicate the experts' behaviour on classifying depicted scenery with Scikit-learn python library \citep{scikit-learn}.
For each image, a unique class label was provided, effectively forming a multi-class classification task.
The same 512-dimensional visual representation $\boldsymbol{H}^V$ introduced in Section~\ref{sec:vis feature} was generated from the images as the inputs.
Classifiers including Multi-layer Perceptron (MLP) (shallow neural network) \citep{hinton1990connectionist}, K-Nearest Neighbour (KNN) \citep{altman1992introduction}, Gaussian Naive Bayes (GNB) \citep{rish2001empirical}, Support Vector Machine (SVM) \citep{platt1999probabilistic}, Random Forest (RF) \citep{breiman2001random}, and Bagging Classifier \citep{breiman1996bagging} with SVM core (BC-SVM) were first trained and tuned for optimal hyperparameters using 10-fold cross validation (CV) with grid search \citep{arlot2010survey}.
Then the individually-trained models were put into ensemble-learning settings as both a voting \citep{zhou2012ensemble} and a stacking classifier \citep{breiman1996stacked}.
All trained models were tested on validation and test datasets to evaluate their performance.
Details of the machine learning models are given in Appendix~\ref{sec:App_ml}.

Both ensemble models were further applied in images collected in this study.
Similar to the HV labels described in Section~\ref{sec:value_label}, the label generation step of HA could be formulated as:
\begin{align}
\label{eq:att_label}
    \boldsymbol{y}^\text{VOTE}_i=\boldsymbol{h}_\text{VOTE}(\boldsymbol{h}^V_i|{\Theta}_\text{VOTE}, \mathcal{M}, \boldsymbol{\Theta}_\mathcal{M}),\\
    \boldsymbol{y}^\text{STACK}_i=\boldsymbol{h}_\text{STACK}(\boldsymbol{h}^V_i|{\Theta}_\text{STACK}, \mathcal{M}, \boldsymbol{\Theta}_\mathcal{M}),\\
    \boldsymbol{Y}^\text{HA} := [\boldsymbol{y}^\text{HA}_i]_{9\times K}, \boldsymbol{y}^\text{HA}_{i}= \frac{\boldsymbol{y}^\text{VOTE}_{i}+\boldsymbol{y}^\text{STACK}_{i}}{2}.
\end{align}
where $\boldsymbol{h}_*$ is an ensemble model taking all parameters $\boldsymbol{\Theta}_\mathcal{M}$ from each ML model in set $\mathcal{M}$; and $\boldsymbol{y}^*_i$ is a 9-dimensional logit vector as soft-label predictions. Similarly, the confidence of models for top-$n$ prediction is:
\begin{align}
\label{eq:att_conf}
    \boldsymbol{K}^\text{HA} = [\boldsymbol{\kappa}^\text{HA}_i]_{2\times K},
    \boldsymbol{\kappa}^\text{HA}_i = [\kappa^{\text{HA}(0)}_i,\kappa^{\text{HA}(1)}_i]^T,\\
    \kappa^{\text{HA}(0)}_i= \sum_{n_0=1}^{n}\frac{\text{max}(\boldsymbol{y}^\text{VOTE}_{i},n_0)+\text{max}(\boldsymbol{y}^\text{STACK}_{i},n_0)}{2},\\
    \kappa^{\text{HA}(1)}_i=\text{IoU}(\text{argmx}(\boldsymbol{y}^\text{VOTE}_{i},n), \text{argmx}(\boldsymbol{y}^\text{STACK}_{i},n)).
\end{align}
This confidence indicator matrix $\boldsymbol{K}^\text{HA}$ could also be the filter for heritage attributes labels $\boldsymbol{Y}^\text{HA}$.

\subsection{Multi-Graph Construction}\label{sec:graph construction}
Three types of similarities/ relations among posts were considered to compose the links connecting the post nodes: \emph{temporal similarity} ({posts with images taken during the same time period}), \emph{social similarity} ({posts owned by the same people, by friends, and by people who share mutual interests}), and \emph{spatial similarity} ({posts with images taken at the same or nearby locations}).
All three could be deduced from the contextual information in Section~\ref{sec:cntx feature}.
As the result, an undirected weighted multi-graph (also known as Multi-dimensional Graph in \cite{ma2021deep}) with the same node set and three different link sets could be constructed as $\mathcal{G}=(\mathcal{V}, \{\mathcal{E}^\text{TEM},\mathcal{E}^\text{SOC},\mathcal{E}^\text{SPA}\}, \{\boldsymbol{w}^\text{TEM},\boldsymbol{w}^\text{SOC},\boldsymbol{w}^\text{SPA}\})$, where $\mathcal{V}=\{v_1, v_2, ..., v_K\}$ is the node set of all the posts, $\mathcal{E}^{(*)}\subseteq \mathcal{V}\times \mathcal{V}$ is the set of all links connecting two posts of one similarity type, and the weight vector $\boldsymbol{w}^{(*)}:=[w^{(*)}_e]_{|\mathcal{E}^{(*)}| \times 1} \in \mathbb{R}_+^{|\mathcal{E}^{(*)}|}$ 
marks the strength of connections.
The multi-graph $\mathcal{G}$ could also be easily split into three simple undirected weighted graphs $\mathcal{G}^\text{TEM}=(\mathcal{V}, \mathcal{E}^\text{TEM}, \boldsymbol{w}^\text{TEM})$, $\mathcal{G}^\text{SOC}=(\mathcal{V}, \mathcal{E}^\text{SOC}, \boldsymbol{w}^\text{SOC})$, and $\mathcal{G}^\text{SPA}=(\mathcal{V}, \mathcal{E}^\text{SPA}, \boldsymbol{w}^\text{SPA})$ concerning each type of similarities.
Each of $\mathcal{G}^*$ corresponds to a weighted adjacency matrix $\boldsymbol{A}^{(*)}:=[a^{(*)}_{i,i'}]_{K\times K}\in \mathbb{R}_+^{K\times K}, i,i' \in [1,K]$, such that:
\begin{equation}
    \label{eq:weighted adj}
    a^{(*)}_{i,i'}=
    \begin{cases}
      {w}^{(*)}_e & \text{if the }e_{\text{th}} \text{ element of } \mathcal{E} \text{ is }(v_{i},v_{i'}),\\
      0 & \text{otherwise}.
    \end{cases}
\end{equation}

The three weighted adjacency matrices could be respectively obtained as follows:

\paragraph{Temporal Links}
Let $\boldsymbol{\mathfrak{T}}_{|\mathcal{T}|\times |\mathcal{T}|}$ denote a symmetric tridiagonal matrix where the diagonal entries are all 1 and off-diagonal non-zero entries are all $\alpha_{\mathcal{T}}$, where $\alpha_{\mathcal{T}} \in [0,1)$ is a parametric scalar:
\begin{equation}
    \label{eq:temporal_tridiagonal}
    \boldsymbol{\mathfrak{T}}_{|\mathcal{T}|\times |\mathcal{T}|}:=
    \begin{pmatrix}
1 & \alpha_{\mathcal{T}} & 0 & \cdots & 0 & 0\\
\alpha_{\mathcal{T}} & 1 & \alpha_{\mathcal{T}} & \cdots & 0 & 0\\
0 & \alpha_{\mathcal{T}} & 1 & \cdots & 0 & 0\\
\vdots  & \vdots &\vdots & \ddots & \vdots &\vdots \\
0 & 0 & 0 & \cdots & 1 & \alpha_{\mathcal{T}}\\
0 & 0 & 0 & \cdots & \alpha_{\mathcal{T}} & 1 
\end{pmatrix},
\end{equation}
then the weighted adjacency matrix $\boldsymbol{A}^{\text{TEM}}_{K \times K}$ for temporal links could be formulated as:
\begin{equation}
    \label{eq:adj_TEM}
    \boldsymbol{A}^{\text{TEM}} = \boldsymbol{T}^\mathsf{T} \boldsymbol{\mathfrak{T}}\boldsymbol{T}, \boldsymbol{A}^{\text{TEM}}\in \{0,\alpha_{\mathcal{T}},1\}^{K \times K},
\end{equation}
where $\boldsymbol{T}_{|\mathcal{T}| \times K}$ is the one-hot embedding of timestamp for posts mentioned in Equation~\ref{eq:One-hot T}.
For simplicity, $\alpha_{\mathcal{T}}$ is set to 0.5.
With such a construction, all the posts from which the images were originally taken in the same week would have a weight of $w^\text{TEM}_e=1$ connecting them in $\mathcal{G}^\text{TEM}$, and posts with images taken in nearby weeks in a chronological order would have a weight of $w^\text{TEM}_{e'}=0.5$.

\paragraph{Social Links}
Let $\boldsymbol{\mathfrak{U}}_{|\mathcal{U}|\times |\mathcal{U}|}$ denote a symmetric matrix as a linear combination of three matrices marking the social relations among the users: 
\begin{equation}
    \label{eq:social_relation}
    \boldsymbol{\mathfrak{U}}_{|\mathcal{U}|\times |\mathcal{U}|} = \frac{\alpha^{(1)}_{\mathcal{U}} \boldsymbol{I} + \alpha^{(2)}_{\mathcal{U}} \boldsymbol{A}^{\mathcal{U}} + \alpha^{(3)}_{\mathcal{U}} (\boldsymbol{A}^{\mathcal{U'}}>\beta_{\mathcal{U}})}{\alpha^{(1)}_{\mathcal{U}}+\alpha^{(2)}_{\mathcal{U}}+\alpha^{(3)}_{\mathcal{U}}},
\end{equation}
where $\boldsymbol{I} \in \{0,1\}^{|\mathcal{U}|\times |\mathcal{U}|}$ is a diagonal matrix of $1$s for the \emph{self relation}, $\boldsymbol{A}^{\mathcal{U}} \in \{0,1\}^{|\mathcal{U}|\times |\mathcal{U}|}$ is the matrix mentioned in Equation~\ref{eq:friendship} for the \emph{friendship relation}, $(\boldsymbol{A}^{\mathcal{U'}}>\beta_{\mathcal{U}})\in \{0,1\}^{|\mathcal{U}|\times |\mathcal{U}|}$ is a mask on the matrix $\boldsymbol{A}^{\mathcal{U'}}$ introduced in Equation~\ref{eq:interest} for the \emph{common-interest relation} above a certain threshold $\beta_{\mathcal{U}}\in (0,1)$, and $\alpha^{(1)}_{\mathcal{U}},\alpha^{(2)}_{\mathcal{U}},\alpha^{(3)}_{\mathcal{U}}\in \mathbb{R}_+$ are parametric scalars to balance the weights of different social relations.
The weighted adjacency matrix $\boldsymbol{A}^{\text{SOC}}_{K \times K}$ for social links could be formulated as:
\begin{equation}
    \label{eq:adj_SOC}
    \boldsymbol{A}^{\text{SOC}} = \boldsymbol{U}^\mathsf{T} \boldsymbol{\mathfrak{U}}\boldsymbol{U}, \boldsymbol{A}^{\text{SOC}}\in [0,1]^{K \times K},
\end{equation}
where $\boldsymbol{U}_{|\mathcal{U}| \times K}$ is the one-hot embedding of owner/user for posts mentioned in Equation~\ref{eq:One-hot U}.
For simplicity, the threshold $\beta_{\mathcal{U}}$ is set to 0.05 and the scalars $\alpha^{(1)}_{\mathcal{U}},\alpha^{(2)}_{\mathcal{U}},\alpha^{(3)}_{\mathcal{U}}$ are all set to 1.
With such a construction, all posts uploaded by the same user would have a weight of $w^\text{SOC}_e=1$ connecting them in $\mathcal{G}^\text{SOC}$, posts by friends with common interests (of more than 5\% common groups subscriptions) would have a weight of $w^\text{SOC}_{e'}=\frac{2}{3}$, and posts by either friends with little common interests or strangers with common interests would have a weight of $w^\text{SOC}_{e''}=\frac{1}{3}$.

\paragraph{Spatial Links}
Let $\boldsymbol{\mathfrak{S}}:=[\mathfrak{s}_{l,l'}]\in [0,1]^{|V| \times |V|}, l, l' \in [1, |V|]$ denote a symmetric matrix computed with simple rules showing the spatial closeness (conductance) of nodes from the spatial graph $G=(V,E,\boldsymbol{w})$ mentioned in Section~\ref{sec:cntx feature}, whose weights $\boldsymbol{w}:=[w_e]_{|E|\times 1} \in [0,20.0]^{|E|}$ originally showed the distance of nodes (resistance):
\begin{equation}
    \label{eq:spatial closeness}
    \mathfrak{s}_{l,l'}=
    \begin{cases}
      \frac{20-w_e}{20} & \text{if the }e_{\text{th}} \text{ element of } E \text{ is }(\upsilon_{l},\upsilon_{l'}),\\
      0 & \text{otherwise}.
    \end{cases}
\end{equation}
The weighted adjacency matrix $\boldsymbol{A}^{\text{SPA}}_{K \times K}$ for spatial links could be formulated as:
\begin{equation}
    \label{eq:adj_SPA}
    \boldsymbol{A}^{\text{SPA}} = \boldsymbol{S}^\mathsf{T} \boldsymbol{\mathfrak{S}}\boldsymbol{S}, \boldsymbol{A}^{\text{SPA}}\in [0,1]^{K \times K},
\end{equation}
where $\boldsymbol{S}_{|V| \times K}$ is the one-hot embedding of spatial location for posts mentioned in Equation~\ref{eq:One-hot Geo}.
With such a construction, posts located at the same spatial node would have a weight of $w^\text{SPA}_e=1$ in $\mathcal{G}^\text{SPA}$, and posts from nearby spatial nodes would have a weight linearly decayed based on distance within a maximum transport time of 20 minutes.

Additionally, the multi-graph $\mathcal{G}$ could be simplified as a simple composed graph $\mathcal{G}'=(\mathcal{V},\mathcal{E}')$ with a binary adjacency matrix $\boldsymbol{A} \in \{0,1\}^{K\times K}$, such that:
\begin{equation}
    \label{eq:simple graph}
    \boldsymbol{A} := (\boldsymbol{A}^{\text{TEM}}>0) \vee (\boldsymbol{A}^{\text{SOC}}>0) \vee (\boldsymbol{A}^{\text{SPA}}>0),
\end{equation}
which connects two nodes of posts if they are connected and similar in at least one contextual relationship.

All graphs were constructed with NetworkX python library \citep{hagberg2008exploring}.
The rationale under constructing various graphs has been briefly described in Section~\ref{sec:Introduction}: the posts close to each other (in temporal, social, or spatial senses) could be arguably similar in their contents, and therefore, also similar in the heritage values and attributes they might convey.
Instead of regarding these similarities as redundancy and e.g., removing duplicated posts by the same user to avoid biasing the analysis, such as in \cite{Ginzarly2019}, this study intends to take advantage of as much available data as possible, since similar posts may enhance and strengthen the information, compensating the redundancies and/or nuances using back-end graph structures.
At later stage of the analysis, the graph of posts could be even coarsened with clustering and graph partitioning methods \citep{karypis1995analysis,lafon2006diffusion,gao2019graph,ma2021deep}, to give an effective summary of possibly similar posts.

\section{Analyses}\label{sec:Analyses}

\subsection{Sample-Level Analyses of Datasets}\label{sec: Quality}
\subsubsection{Generated Visual and Textual Features}\label{sec:consistency}

Table~\ref{T_consistence_vis} shows the consistency of generated visual and textual features.
The visual features compared the scene and attribute predictions on images of difference sizes (150$\times$150 and 320$\times$240 px); and the textual features compared the OUV selection criteria with aggregated (averaged) sentence-level predictions on each sentence from set $\{\mathcal{s}^{(1)}_i, \mathcal{s}^{(2)}_i, ..., \mathcal{s}^{(|\mathcal{S}_i|)}_i\}$ and paragraph-/post-level predictions on set $\mathcal{S}_i$.

\begin{table}[ht]
\scriptsize\centering
\caption{\footnotesize The consistency (the mean and standard deviation of top-$n$ IoU Jaccard Index on predicted sets) of generated features. The best scores for each feature are in bold, and the selected ones for future tasks are underlined.}\label{T_consistence_vis}
\sf\begin{tabular}{lccc}
\toprule
Sets to calculate IoU Jaccard Index & AMS & SUZ & VEN\\
\midrule
\#Compared Posts w. Visual Features & 3727 & 3137 & 2951\\
Top-1 scene predictions & \textbf{.656} & \textbf{.676} & \textbf{.704}\\
--- $\text{argmx}(\boldsymbol{l}^\text{s},1)$ & (.475) & (.468) & (.456)\\
\textbf{Top-5 scene predictions} & {\underline{.615}} & \underline{.636} & \underline{.635}\\
--- $\text{argmx}(\boldsymbol{l}^\text{s},5)$ & (\textbf{\underline{.179}}) & (\textbf{\underline{.238}}) & (\textbf{\underline{.229}})\\
Top-1 attribute predictions & \textbf{.867} & \textbf{.853} & \textbf{.838}\\
--- $\text{argmx}(\boldsymbol{l}^\text{a},1)$ & (.339) & (.354) & (.368)\\
\textbf{Top-10 attribute predictions} & \underline{.820} & \underline{.802} & \underline{.819}\\
--- $\text{argmx}(\boldsymbol{l}^\text{a},10)$ & (\textbf{\underline{.140}}) & (\textbf{\underline{.144}}) & (\textbf{\underline{.139}})\\
\midrule
\#Compared Posts w. Textual Features & 2904 & 754 & 1761\\
Top-1 OUV predictions & .775 & .923 & .714\\
--- $\text{argmx}(\boldsymbol{y}^\text{BERT},1)$ & (.418) & (.267) & (.452)\\
\textbf{Top-3 OUV predictions} & \textbf{\underline{.840}} & \textbf{\underline{.938}} & \textbf{\underline{.791}}\\
--- $\text{argmx}(\boldsymbol{y}^\text{BERT},3)$ & (\textbf{\underline{.246}}) & (\textbf{\underline{.182}}) & (\textbf{\underline{.266}})\\
\bottomrule
\end{tabular}
\end{table}

For both scene and attribute predictions, the means of top-$1$ Jaccard index were always higher than that of top-$n$, however, the smaller variance proved the necessity of using top-$n$ prediction as features.
Note the attribute prediction was more stable than the scene prediction when the image shape changed, this is probably because the attributes usually describe low-level features which could appear in multiple parts in the image, while some critical information to judge the image scene may be lost during cropping and resizing in the original ResNet-18 model.
Considering the relatively high consistency of model performance and the storage cost of images when the dataset would ultimately scale up (e.g., VEN-XL), the following analyses would only be performed on smaller square images of 150$\times$150 px.

The high Jaccard index of OUV predictions showed that averaging the textual features derived from sub-sentences of a paragraph would yield a similar performance of directly feeding the whole paragraph into models, especially when the top-$3$ predictions are of main interests.
Note the higher consistency in Suzhou was mainly a result of the higher proportion of posts only consisting one sentence.

\begin{table}[ht]
\scriptsize\centering
\caption{\footnotesize Descriptive statistics (mean and standard deviation or counts, respectively) of the facial recognition results $\boldsymbol{F}$ as visual features and original language $\boldsymbol{O}$ as textual features.}\label{T_describe_vis}
\sf\begin{tabular}{lcccc}
\toprule
Features & AMS & SUZ & VEN & VEN-XL\\
\midrule
\#Posts w. Faces & 667 & 303 & 166 & 9287\\
\#Faces detected & 1.547 & 1.403 & 1.349 & 1.298\\
--- $\boldsymbol{\mathfrak{f}}_1$ & (.830) & (.707) & (.785) & (.651)\\
Model Confidence & .955 & .956 & .930 & .948\\
--- $\boldsymbol{\mathfrak{f}}_2$ & (.079) & (.081) & (.099) & (.081)\\
Area proportion of faces & .049 & .057 & .077 & .076\\
--- $\boldsymbol{\mathfrak{f}}_3$ & (.112) & (.073) & (.185) & (.112)\\
\midrule
\#Posts w. Texts* & 2904 & 754 & 1761 & 49,823\\
\#Posts in English $\boldsymbol{\mathfrak{o}}_1$ & 1488 & 368 & 640 & 20,271\\
\#Posts in Native Lang $\boldsymbol{\mathfrak{o}}_2$ & 1773 & 27 & 1215 & 28,633\\
\#Posts in Other Lang $\boldsymbol{\mathfrak{o}}_3$ & 536 & 413 & 657 & 21,916\\
\bottomrule
\multicolumn{5}{l}{\scriptsize *Note this is smaller than the sum of the three below, since each}\\
\multicolumn{5}{l}{\scriptsize post can be written in multiple languages.}\\
\end{tabular}
\end{table}

Table~\ref{T_describe_vis} gives descriptive statistics of results that were not compared against different scenarios as in Table~\ref{T_consistence_vis}.
Only a small portion of posts had detected faces on them.
While Amsterdam has the highest proportion of face pictures (17.9\%), Venice has larger average area of faces on the picture (i.e., more selfies and tourist pictures).
These numbers are also assumed to help associate a post to human-activity-related heritage values and attributes.
Considering the languages of the posts, Amsterdam showed a balance between Dutch-speaking locals and English-speaking tourists, Venice showed a balance between Italian-speaking people and non-Italian-speaking tourists, while Suzhou showed a lack of Chinese posts.
This is consistent with the popularity of Flickr as social media in different countries, which also implies that data from other social media could compensate this unbalance if the provisional research questions would be sensitive to the nuance between local and tourist narratives.

\subsubsection{Pseudo-Labels for Heritage Values and Attributes}\label{sec:pseudo-labels}

\begin{figure*}[ht]
\centering
\includegraphics[width=\linewidth]{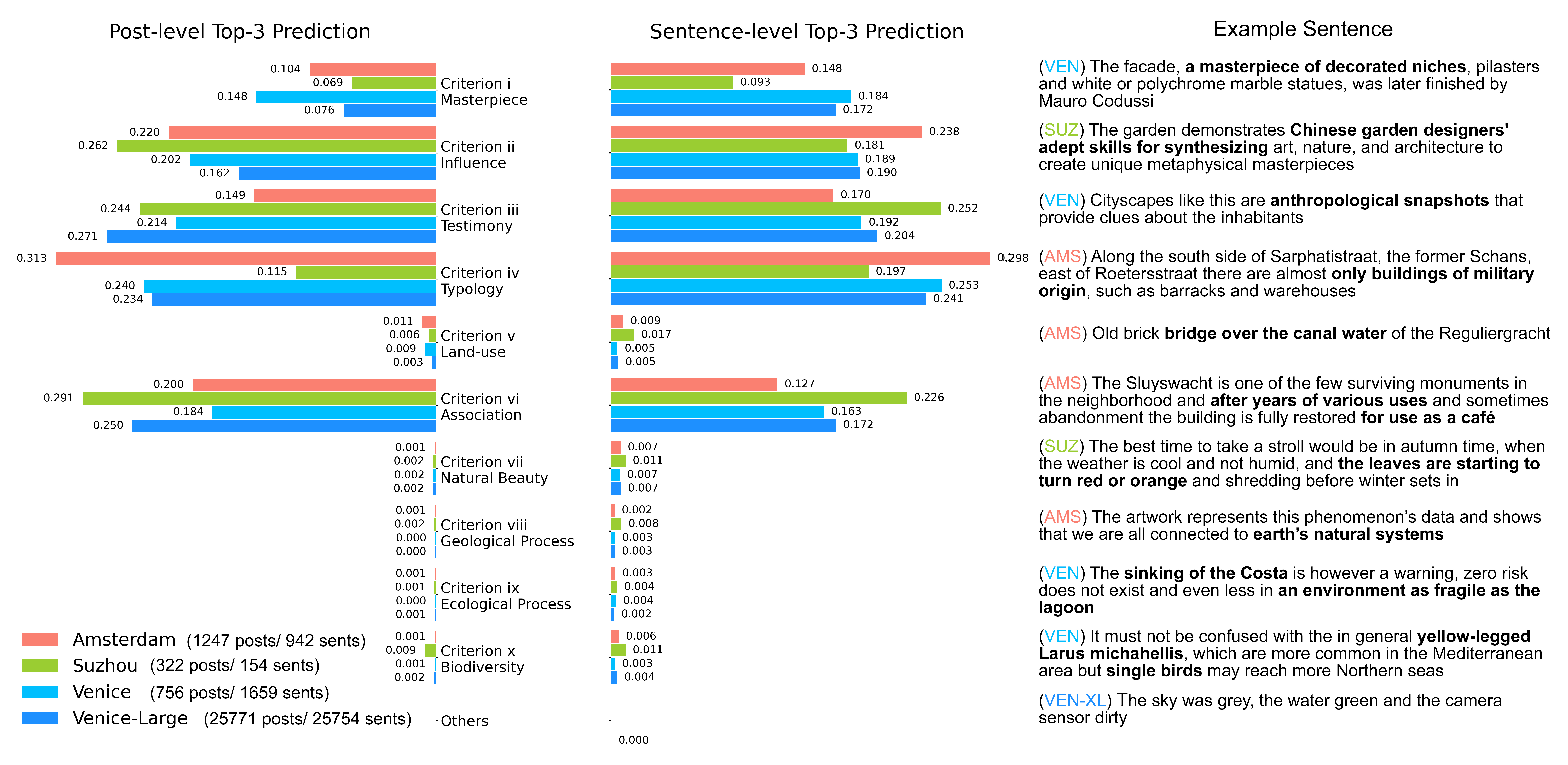}
\caption{\footnotesize The proportion of posts and sentences that are predicted and labeled as each heritage value (OUV selection criterion) as top-3 predictions by both BERT and ULMFiT. One typical sentence from each category is also given.}\label{fig:Values}
\end{figure*}

\begin{figure*}[ht]
\centering
\includegraphics[width=0.9\linewidth]{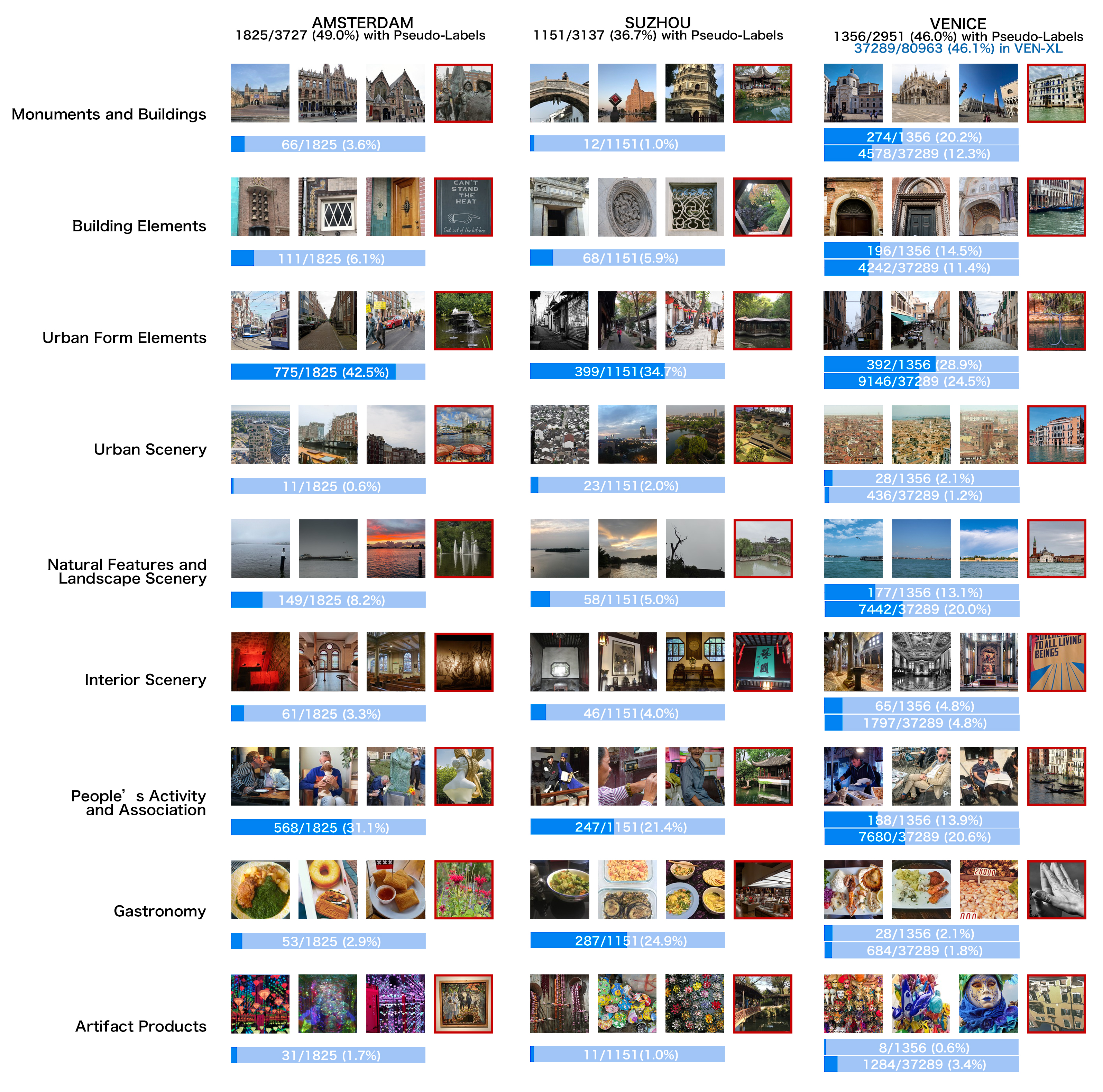}
\caption{\footnotesize Typical image examples in each city labelled as each heritage attribute category (depicted scene) and bar plots of their proportions in the datasets (length of bright blue background bars represent 50\%). Three examples with high confidence and one negative example with low confidence (in red frame) are given. All images are 150$\times$150 px `thumbnails' flagged as `downloadable'.}\label{fig:Attributes}
\end{figure*}

As argued in Section~\ref{sec:value_label}, the label generation process of this paper did not involve human annotators.
Instead, it used thoroughly trained ML models as machine replica of annotators and considered their confidences and agreements as a filter to keep the `high-quality' labels as pseudo-labels.
Similar operations could be found in semi-supervised learning \citep{zhou2010semi, lee2013pseudo, sohn2020fixmatch}.

For heritage values, an average top-3 confidence of $\kappa^\text{HV(0)}>0.75$ and top-3 agreement (Jaccard Index) of $\kappa^\text{HV(1)}>0.5$ was used as the filter for $\boldsymbol{Y}^\text{HV}$.
This resulted around 40-50\% samples with textual data in each city as `labelled', and the rest as `unlabelled'.
Figure~\ref{fig:Values} demonstrates the distribution of `labelled' data about heritage values in each city.
For all cities, cultural values are far more frequent than natural values, consistent with their status of cultural WH.
However, elements related to natural values could still be found and were mostly relevant.
The actual OUV inscribed in WHL mentioned in Table~\ref{T_Case_studies} could all be observed as significantly present (e.g., criteria (i),(ii),(iv) for Amsterdam) except for criterion (v) in Venice and Suzhou, which might be caused by the relatively fewer examples and poorer class-level performance of criterion (v) in the original paper.
Remarkably, criterion (iii) in Amsterdam and criterion (vi) in Amsterdam and Suzhou were not officially inscribed, but appeared to be relevant inducing from social media, inviting further heritage-specific investigations.
The distributions of Venice and Venice-large were more similar in sentence-level predictions (Kullback-Leibler Divergence $D_\text{KL}=.002$, Chi-square $\chi^2=39.515$) than post-level ($D_\text{KL}=.051, \chi^2=518.895$), which might be caused by the specific set of posts sub-sampled in the smaller dataset.

\begin{table}[ht]
\scriptsize\centering
\caption{\footnotesize The performance of models during the cross validation (CV) parameter selection
, on the validation set, and on the test set of data from Tripoli. The best two models for each performance are bold, and the best one underlined.}\label{T_ML}
\sf\begin{tabular}{lrrrrr}
\toprule
ML Model & CV Acc & Val Acc & Val F1 & Test Acc & Test F1\\
\midrule
MLP & .767 & .749 & .70 & .789 & .72\\
KNN & .756 & .724 & .67 & .767 & .71\\
GNB & .738 & .749 & .71 & .800 & .77\\
SVM & \textbf{\underline{.797}} & .754 & .71 & .822 & .78\\
RF & .766 & .734 & .68 & .789 & .72\\
BC-SVM & .780 & .759 & .71 & .811 & .74\\
\midrule
\textbf{VOTE} & .788 & \textbf{.764} & \textbf{\underline{.72}} & \textbf{\underline{.855}} & \textbf{\underline{.82}}\\
\textbf{STACK} & \textbf{.794} & \textbf{\underline{.768}} & \textbf{\underline{.72}} & \textbf{.844} & \textbf{.81}\\
\bottomrule
\end{tabular}
\end{table}

For heritage attributes, Table~\ref{T_ML} shows the performance of ML models mentioned in Section~\ref{sec:attr_label}.
The two ensemble models with voting and stacking settings performed equally well and significantly better than other models (except for CV accuracy of SVM), proving the rationale of using both classifiers for heritage attribute label prediction.
An average top-1 confidence of $\kappa^\text{HA(0)}>0.7$ and top-1 agreement of $\kappa^\text{HA(1)}=1$ was used as the filter for $\boldsymbol{Y}^\text{HA}$.
This filter resulted around 35-50\% images in each city as `labelled', and the rest as `unlabelled'.
Figure~\ref{fig:Attributes} demonstrates the distribution of `labelled' data about heritage attributes 
in each city.
It is remarkable that although the models were only trained on data from Tripoli, they performed reasonably well in unseen cases of Amsterdam, Suzhou, and Venice, capturing typical scenes of monumental buildings, architectural elements, and gastronomy etc., respectively.
Although half of the collected images were treated as `unlabelled' due to low confidence, the negative examples are not necessarily incorrect (e.g., with \emph{Monuments and Buildings}).
For all cities, \emph{Urban Form Elements} and \emph{People's Activity and Association} are the most dominant classes, consistent with the fact that the most Flickr images are taken on the streets.
Seen from the bar plots in Figure~\ref{fig:Attributes}, the classes were relatively unbalanced, suggesting that more images from small classes might be needed or at least augmented in future applications.
Furthermore, the distributions of Venice and Venice-large are similar to each other ($D_\text{KL}=.076, \chi^2=188.241$), suggesting a good representativeness of the sampled small dataset.

\begin{figure*}[ht]
\centering
\includegraphics[width=0.9\linewidth]{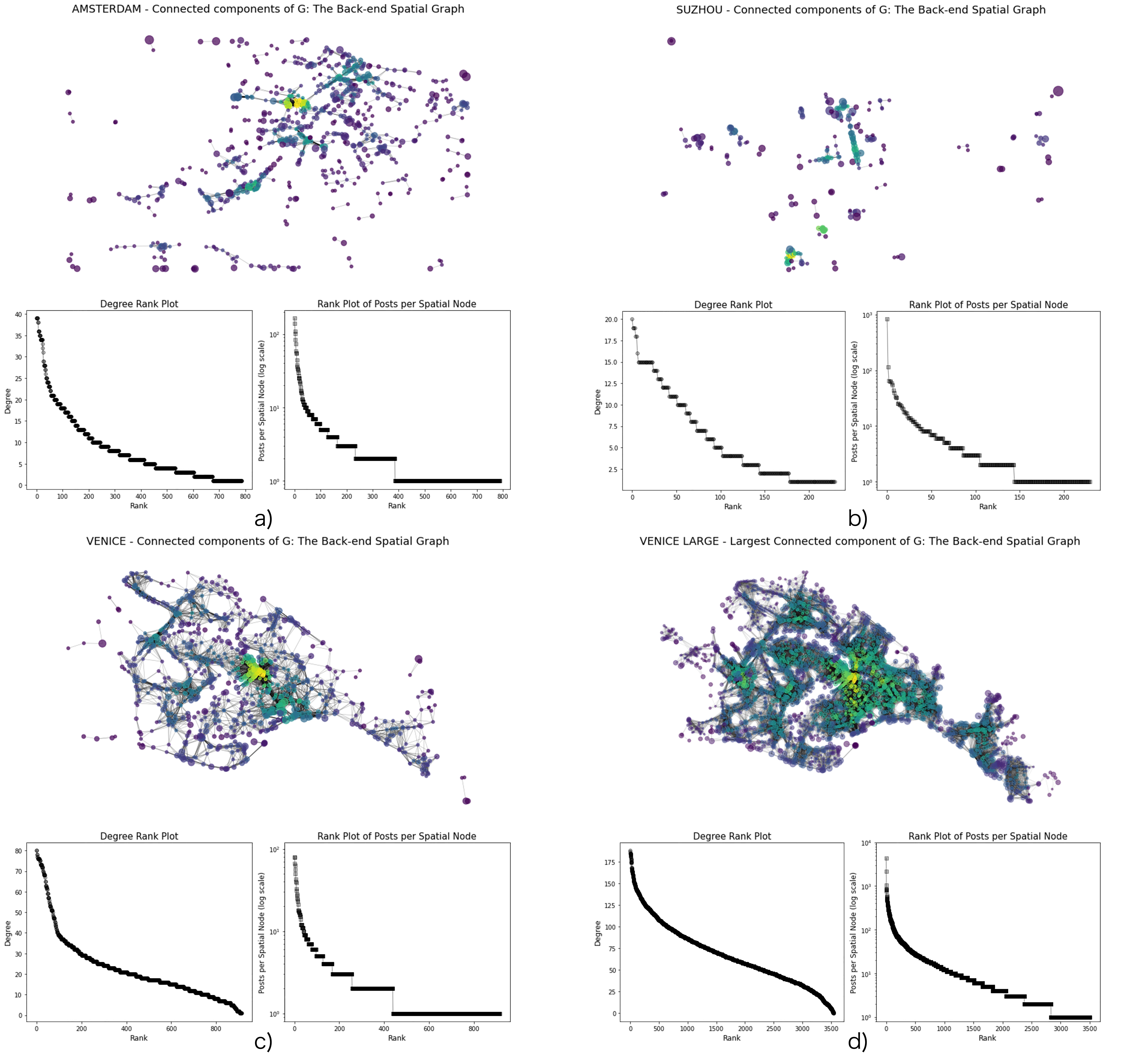}
\caption{\footnotesize The back-end geographical networks for three case studies, respectively showing the graph structure, degree ranking distribution, and the ranking distribution of posts per geo-spatial node (on a logarithm scale) in Amsterdam, Suzhou, Venice, and Venice-XL. The sizes of nodes denote the number of nearby posts allocated to the nodes, and the colors of nodes illustrate the degree of the node on the graph. Each link connects two nodes reachable to each other within 20 minutes.}\label{fig:GEO_structure}
\end{figure*}

\subsection{Graph-Level Analyses of Datasets}\label{sec: Analysis}
\subsubsection{Back-end Geographical Network}
The back-end spatial structures of post locations as graphs $G=(V,E,\boldsymbol{w})$ were visualized in Figure~\ref{fig:GEO_structure}.
Further graph statistics in all cities were given in Table~\ref{T_geo}.
The urban fabric is more visible in Venice than the other two cities, as there is always a dominant large component connecting most nodes in the graph, leaving fewer unconnected isolated nodes alone.
While in Amsterdam, more smaller connected components exist together with a large one, and in Suzhou, the graph is even more fragmented with smaller components.
This is possibly related to the distribution of tourism destinations, which is also consistent with the zoning typology of WH property concerning urban morphology \citep{PereiraRoders2010, Valese2020}: for Venice, the Venetian islands are included together with a larger surrounding lagoon in the WH property (formerly referred to as core zone), and are generally regarded as a tourism destination as a whole; for Amsterdam, the WH property is only a part of the old city being mapped where tourists could freely wander and take photos in areas not listed yet interesting as tourism destinations; while for Suzhou, the WH properties are themselves fragmented gardens distributed in the old city, also being the main destinations visited by (foreign) tourists.

\begin{table}[ht]
\scriptsize\centering
\caption{\footnotesize The statistics for the back-end Geographical Network $G=(V,E,\boldsymbol{w})$.}\label{T_geo}
\sf\begin{tabular}{lrrrr}
\toprule
Graph Features & AMS & SUZ & VEN & VEN-XL\\
\midrule
\#Nodes in $V$ & 788 & 230 & 915 & 3549\\
\#Edges in $E$ & 3331 & 680 & 10,385 & 120,033\\
\#Connected Components & 72 & 38 & 6 & 13\\
\#Nodes Largest CC* & 355 & 50 & 897 & 3498\\
Graph Density & .011 & .026 & .025 & .019\\
\#Isolated Nodes in $V_0\backslash V$ & 157 & 88 & 20 & 22\\
\bottomrule
\multicolumn{5}{l}{\scriptsize *Connected Components.}\\
\end{tabular}
\end{table}

Furthermore, the two types of rank-size plots showing respectively the degree distribution and the posts-per-node distribution showed similar patterns, the latter being more heavy-tailed, a typical characteristic of large-scale complex networks \citep{barabasi2013network,eom2015tail}, while the back-end spatial networks are relatively more regular.

\subsubsection{Multi-Graphs}

\begin{figure*}[ht]
\centering
\includegraphics[width=0.95\linewidth]{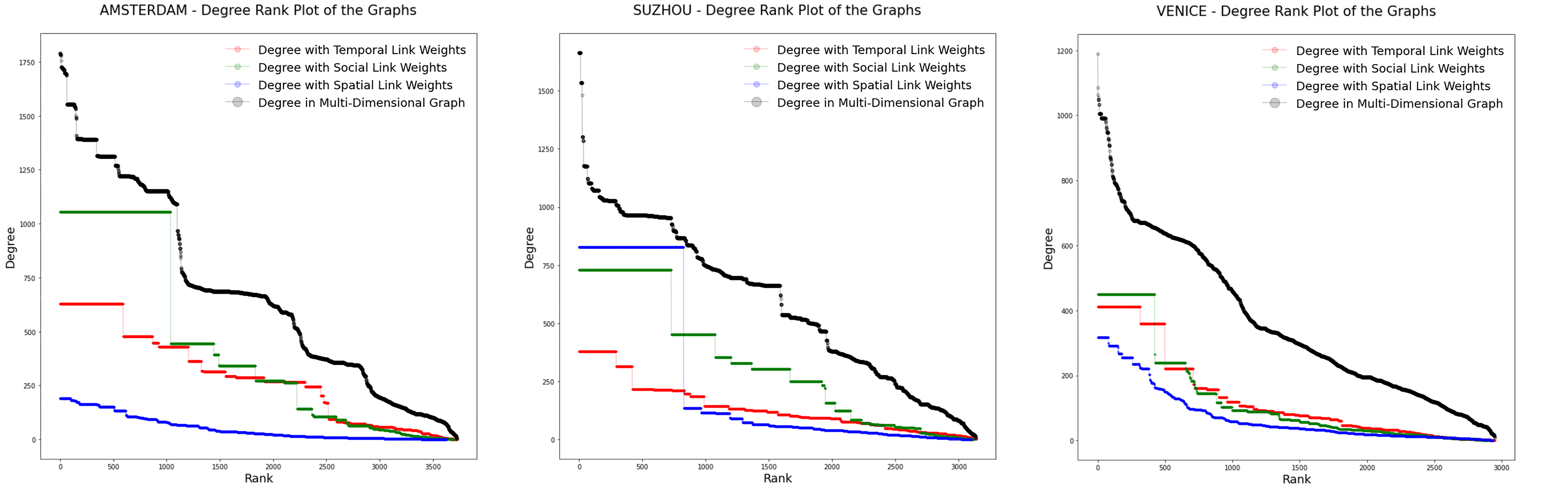}
\caption{\footnotesize The rank-size plots of the degree distributions in the three cases of Amsterdam, Suzhou, and Venice, with regard to the temporal links, social links, spatial links, as well as the entire multi-graph.}\label{fig:degree_distribution}
\end{figure*}

Table~\ref{T_graph_size} shows graph statistics of three constructed sub-graphs $\mathcal{G}^\text{TEM},\mathcal{G}^\text{SOC},\mathcal{G}^\text{SPA}$ with different link types within the multi-graph $\mathcal{G}$, and the simple composed graph $\mathcal{G}'$ for each city, while Figure~\ref{fig:degree_distribution} plots their [weighted] degree distributions, respectively.
The multi-graphs are further visualized in Appendix Figure~\ref{fig:graph_vis}.

\begin{table}[ht]
\scriptsize\centering
\caption{\footnotesize The statistics for the multi-graphs.}\label{T_graph_size}
\sf\begin{tabular}{lrrr}
\toprule
Graph Features & AMS & SUZ & VEN\\
\midrule
\multicolumn{4}{l}{Temporal Graph $\mathcal{G}^\text{TEM}=(\mathcal{V}, \mathcal{E}^\text{TEM},\boldsymbol{w}^\text{TEM})$}\\
\#Nodes* & 3727 & 3137 & 2951\\
\#Edges & 692,839 & 293,328 & 249,120\\
Diameter & 145 & 116 & 270\\
Graph Density & .100 & .060 & .057\\
\midrule
\multicolumn{4}{l}{Social Graph $\mathcal{G}^\text{SOC}=(\mathcal{V}, \mathcal{E}^\text{SOC},\boldsymbol{w}^\text{SOC})$}\\
\#Nodes** & 3696 & 3120 & 2916\\
\#Edges & 877,584 & 602,821 & 242,576\\
\#Connected Components & 47 & 56 & 60\\
\#Nodes Largest CC & 2694 & 942 & 2309\\
Diameter Largest CC & 7 & 6 & 10\\
Graph Density & .129 & .124 & .057\\
\midrule
\multicolumn{4}{l}{Spatial Graph $\mathcal{G}^\text{SPA}=(\mathcal{V}, \mathcal{E}^\text{SPA},\boldsymbol{w}^\text{SPA})$}\\
\#Nodes** & 3632 & 3102 & 2938\\
\#Edges & 135,079 & 415,049 & 221,414\\
\#Connected Components & 134 & 91 & 13\\
\#Nodes Largest CC & 1485 & 829 & 2309\\
Diameter Largest CC & 22 & 1 & 22\\
Graph Density & .020 & .086 & .051\\
\midrule
\multicolumn{4}{l}{Simple Composed Graph $\mathcal{G}'=(\mathcal{V}, \mathcal{E}')$}\\
\#Nodes* & 3727 & 3137 & 2951\\
\#Edges & 1,271,171 & 916,496 & 534,513\\
Diameter & 4 & 5 & 4\\
Graph Density & .183 & .186 & .123\\
\bottomrule
\multicolumn{4}{l}{\scriptsize *By definition a connected graph (only one connected component).}\\
\multicolumn{4}{l}{\scriptsize **The isolated nodes with no links are not counted here.}\\
\end{tabular}
\end{table}

The three link types provided heterogeneous characteristics:
1) the temporal graph is by definition connected, where the highest density in Amsterdam suggested the largest number of photos taken in consecutive times, while the largest diameter in Venice suggested the broadest span of time;
2) the social graph is structured by the relationship of users, where the largest connected components showed clusters of posts shared either by the same user, or by users who are friends or with mutual interests, the size of which in Suzhou is small because of the fewest users shown in Table~\ref{T_Case_studies};
3) the spatial graph shows similar connectivity pattern with the back-end spatial graphs, where the extremely small diameter and the largest density in Suzhou reassured the fragmented positions of posts;
4) although the degree distribution of three sub-graphs fluctuated due to different socio-economic and spatio-temporal characteristics of different cities, that of the simple composed graph showed similar elbow-shape patterns, with similar density and diameter.
Moreover, the heterogeneous graph structures suggest that different parameters and/or backbone models need to be fit and fine-tuned with each link type, a common practice for deep learning on multi-graphs.

\section{Discussion}\label{sec:Discussion}
\subsection{Provisional Tasks for Urban Data Science}\label{sec:urban data science}



\begin{table*}[ht]
\scriptsize\centering
\caption{\label{T_provisional_tasks}
\footnotesize A few provisional tasks with formal problem definitions that could be performed. Potential scientific and social relevance for Machine Learning community, and urban and/or heritage researchers, respectively, are given.}
\sf\begin{tabular}{lp{105pt}p{63pt}p{105pt}p{123pt}}
\toprule
ID & Problem Definition & Type of Task & As an \emph{ML/SNA} Problem & As an \emph{Urban/Heritage Study} Question\\ 
\midrule
0&$\boldsymbol{X}^\text{vis}\mapsto \boldsymbol{Y}^\text{HV}|\boldsymbol{K}^\text{HV}$ & Image Classification \emph{\color{gray} (semi-supervised)} & Using visual features to infer categories induced from (possibly missing) texts with co-training \citep{blum1998combining} in few-shot learning settings \citep{wang2020generalizing}. & As the latest advances in heritage value assessment have been discovering the added value of inspecting texts \citep{TarrafaSilva2010}
, can values also be seen and retrieved from scenes of images? 
\\
1&$\boldsymbol{X}^\text{tex}\mapsto \boldsymbol{Y}^\text{HA}|\boldsymbol{K}^\text{HA}$& Text Classification \emph{\color{gray} (semi-supervised)} & Using textual features to infer categories induced from images possibly with attention mechanisms \citep{vaswani2017attention}. & How to relate the textual descriptions to certain heritage attributes \citep{Gomez2019530}? Are there crucial hints other than appeared nouns?\\
2&$\boldsymbol{X}:=\left\{\boldsymbol{X}^\text{vis},\boldsymbol{X}^\text{tex}\right\}\mapsto \boldsymbol{Y}:=\left\{\boldsymbol{Y}^\text{HV}|\boldsymbol{K}^\text{HV},\boldsymbol{Y}^\text{HA}|\boldsymbol{K}^\text{HA}\right\}$& Multi-modal Classification \emph{\color{gray} (semi-supervised)} & Using multi-modal (multi-view) features to make inference, either with training joint representations or by making early and/or late fusions \citep{blum1998combining,Baltrusaitis2019}. & How can heritage values and attributes be jointly inferred from the combined information of both visual scenes and textual expressions \citep{Ginzarly2019}? How can they complement each other?\\
3&$\boldsymbol{X}, \boldsymbol{A}\mapsto \boldsymbol{Y}$& Node Classification \emph{\color{gray} (semi-supervised)} & Test-beds for different graph filters such as Graph Convolution Networks \citep{kipf2016semi} and Graph Attention Networks \citep{velivckovic2017graph}. & How can the contextual information of a post contribute to the inference of its heritage values and attributes? What is the contribution of time, space, and social relations \citep{miah2017big}?\\
4&$\boldsymbol{X},\boldsymbol{Y}, \boldsymbol{A}\mapsto \boldsymbol{A}+\boldsymbol{A}_\text{new}$& Link Prediction \& Recommendation System \emph{\color{gray} (semi-supervised)} & Test-beds for link prediction algorithms \citep{adamic2003friends} considering current graph structure and node features. What is the probability that other links also should exist? & Considering the similarity of posts, would there be heritage values and attributes that also suit the interest of another user, fit another location, and/or reflect another period of time \citep{majid2013context}?\\
5&$\boldsymbol{X},\boldsymbol{Y}, \boldsymbol{A}\mapsto \hat{\boldsymbol{X}},\hat{\boldsymbol{Y}},\hat{\boldsymbol{A}}$& Graph Coarsening \emph{\color{gray} (unsupervised)} & Test-beds for graph pooling \citep{ma2021deep} and graph partitioning \citep{karypis1995analysis} algorithms to generate coarsened graphs \citep{pang2021graph} in different resolutions. & How can we summarize, aggregate, and eventually visualize the large-scale information from the social media platforms based on their contents and contextual similarities \citep{cho2022classifying}?\\
6&$\boldsymbol{X},\boldsymbol{Y}, \boldsymbol{A}\mapsto \boldsymbol{y}_{\mathcal{G}}^\text{HV}|\boldsymbol{Y}^\text{HV},\boldsymbol{y}_{\mathcal{G}}^\text{HA}|\boldsymbol{Y}^\text{HA}$& Graph Classification \emph{\color{gray} (supervised)} & Test-beds for graph classification algorithms \citep{zhang2018end} when more similar datasets have been collected and constructed in more case study cities. & Can we summary the social media information of any city with World Heritage property so that the critical heritage values and attributes could be directly inferred \citep{Monteiro2014}?\\
7&$\boldsymbol{X},\boldsymbol{Y}, \boldsymbol{A}\mapsto \boldsymbol{\mathfrak{I}},\boldsymbol{\mathcal{S}}$& Image/Text Generation \emph{\color{gray} (supervised)} & Using multi-modal features to generate the missing and/or unfit images and/or textual descriptions, probably with Generative Adversarial Network \citep{goodfellow2014generative}. & How can a typical image and/or textual description of certain heritage values and attributes at a certain location in a certain time by a certain type of user in a specific case study city be queried or even generated \citep{Gomez2019530}?\\
8&$\boldsymbol{X},\boldsymbol{Y},\boldsymbol{A}^\text{TEM},\boldsymbol{A}^\text{SOC},\boldsymbol{A}^\text{SPA}\mapsto \boldsymbol{{R}}+\boldsymbol{{R}}^\text{TEM}+\boldsymbol{{R}}^\text{SOC}+\boldsymbol{{R}}^\text{SPA}$& Attributed Multi-Graph Embedding \emph{\color{gray} (self-supervised)} & Respectively generating a universal embedding and a context-specific embedding for each type of links in the multi-dimensional network \citep{ma2018multi}, probably with random walks on graphs. & How are heritage values and attributes distributed and diffused in different contexts? Is the First Law of Geography \citep{tobler1970computer} still valid in the specific social, temporal and spatial graphs?\\
9&$\boldsymbol{X}^{(k)},\boldsymbol{Y}^{(k)}, \boldsymbol{A}^{(k)},\boldsymbol{T}\mapsto \boldsymbol{X}^{(k+1)}, \boldsymbol{Y}^{(k+1)}, \boldsymbol{A}^{(k+1)}$& Dynamic Prediction \emph{\color{gray} (self-supervised)} & Given the current graph structure and its features stamped with time steps, how shall it further evolve in the next time steps \citep{nguyen2018continuous, ren2019deep}? & How are the current expressions of heritage values and attributes in a city influencing the emerging post contents, the tourist behaviours, and the planning decision making \citep{bai2021_2, zhang2020graph}?\\
\bottomrule
\end{tabular}
\end{table*}

The datasets introduced in this paper could be used to answer questions from the perspectives of machine learning and social network analysis as well as heritage and urban studies.
Table~\ref{T_provisional_tasks} gives a few provisional tasks that could be realised using the collected datasets of this paper, and further datasets to be collected using the same introduced workflow.
These problems would use some or all of extracted features (visual, textual, contextual), generated labels (heritage values and attributes), constructed graph structures, and even raw data as input and output components to find the relationship function among them. 
Some problems (such as 0, 1, 2, and 6) were marked in Figure~\ref{fig:DataFlow} as question marks.
Some problems are more interesting as ML/SNA problems (such as 4, 7 and 8), some are more fundamental for urban data science (such as 0, 1 and 6).
While the former tends towards the technical and theoretical end of the whole potential range of the datasets, the latter tends towards the application end.
However, to reach a reasonable performance during applications and discoveries, as is the main concern and interest for urban data science, further technical investigations and validations would be indispensable.

In summary, the core question from collecting such datasets could be formulated as: while heritage values and attributes have been historically inspected from site visiting and document reviewing by experts, can computational methods and/or artificial intelligence help accelerate and aid the process of knowledge documentation and comparative studies by mapping and mining multi-modal social media data? 
Even if acceleration of the processes is not a priority, the provision of such a workflow is aimed to encourage consistency and inclusion of communities in the discourse of cherishing, protecting, and preserving cultural heritage. 
In other words, the machine can represent the voice of the community.

Note a further distinction needs to be made within the extracted heritage values and attributes, as basically they may be clustered into three categories:
1) core heritage values and attributes officially listed and recognized that thoroughly defined the heritage status;
2) values and attributes relevant to conservation and preservation practice;
3) other values and attributes not specifically heritage-related yet being conveyed to the same heritage property by ordinary people.
This distinction should be made clear of for practitioners intending to make planning decisions based on the conclusions drawn from studying such datasets.


\subsection{Limitations and Future Steps}\label{sec:limitations}

No thorough human evaluations and annotations have been performed during the construction of the datasets presented in this paper.
This manuscript provides a way to walk around that step by using only the confidence and [dis-]agreement of presumably well-trained models as the proxy for the more conventional `inter-annotator' agreement to show the quality of datasets and generate [pseudo-]labels \citep{Nowak2010}.
This resembles the idea of using the consistency, confidence, and disagreement to improve the model performance in semi-supervised learning \citep{zhou2010semi,lee2013pseudo,sohn2020fixmatch}.
For the purpose of introducing a general workflow that could generate more graph datasets, it is preferable to exclude humans from the loop as it would be a bottleneck limiting the process, both in time and monetary resources, and in demanded domain knowledge.
However, for applications where more accurate conclusions are needed, human evaluations on the validity, reliability, and coherence of the models are still needed.
It is suggested to inspect some predicted results to have a clear sense of the performance before use.
As the step of [pseudo-]label generation was relatively independent from the other steps introduced in this paper, higher-quality labels annotated and evaluated by experts and/or crowd-workers could still be added at a later stage as augmentation or even replacement, as an active learning process \citep{settles2011theories,prince2004does,zhu2009introduction}.
Moreover, generating labels of heritage values and attributes was only one arbitrary choice to showcase the label generation process.
Yet, it is also possible to apply the same workflow while only replacing the classifiers mentioned in Section~\ref{sec:label_generation} with specific topics related to the interests of future researchers, to answer questions from urban data science and computational social sciences.

While scaling up the dataset construction process, such as from VEN to VEN-XL, a few changes need to be adopted.
For data collection, an updated strategy is described in Appendix~\ref{sec:App_data_collection}.
For feature and label generation, mini-batches and GPU computing significantly accelerated the process.
However, the small graphs from case study cities containing around 3000 nodes already contained edges at scale of millions, making it challenging to scale up in cases such as VEN-XL, the adjacency list of which would be at scale of billions, easily exceeding computer memory.
As a result, VEN-XL has not yet been constructed as a multi-graph.
Further strategies such as using sparse matrices \citep{yuster2005fast} and parallel computing should be considered.
Moreover, the issue of scalability should also be considered for later graph neural network training, since the multi-graphs constructed in this study can get quite dense locally.
Sub-graph sampling methods should be applied to avoid `neighbourhood explosion' \citep{ma2021deep}.

Although the motivation of constructing datasets about heritage values and attributes from social media was to promote inclusive planning processes, the selection of social media platforms already automatically excluded those not using, or even not aware of, the platform, let alone those not using internet.
The scarce usage of Flickr in China, as an example, also suggested that conclusions drawn from such datasets may reflect perspectives from `tourist gaze'\citep{urry2011tourist} rather than local communities, therefore losing some representativeness and generality.
However, the main purpose of this paper is to provide a reproducible workflow with mathematical definitions, not limited to Flickr.
Images and descriptions from other platforms such as Weibo, Dianping, RED, and TikTok that are more popular in China could also add complementary local perspectives.
With careful adaptions, archives, official documents, news articles, academic publications, and interview transcripts could also be constructed in similar formats for fairer comparisons.




\section{Conclusions}\label{sec:Conclusion}

This paper introduced a novel workflow to construct graph-based multi-modal datasets \emph{HeriGraph} concerning heritage values and attributes using data from social media platform Flickr.
State-of-the-art machine learning models have been applied to generate multi-modal features and domain-specific pseudo-labels.
Full mathematical formulation is provided for the feature extraction, label generation, and graph construction processes.
Three case study cities Amsterdam, Suzhou, and Venice containing UNESCO World Heritage properties are tested with the workflow to construct sample datasets, being evaluated and filtered with the consistency of models and qualitative inspections.
Such datasets have the potentials to be applied by both machine learning community and urban data scientists to answer interesting questions with scientific/technical and social relevance, which could also be applied around the globe.

\section*{Acknowledgement}
This study is within the framework of the Heriland-Consortium.
HERILAND is funded by the European Union’s Horizon 2020 research and innovation programme under the Marie Sklodowska-Curie grant agreement No 813883.

\section*{Data Availability Statement}
All the data and codes for this manuscript can be accessed through the following GitHub repository respecting the privacy and copyrights of original post owners: \href{https://github.com/zzbn12345/Heri_Graphs}{https://github.com/zzbn12345/Heri\_Graphs}.

\bibliographystyle{abbrvnat}
\bibliography{VeniceFlickr}  

\begin{thebibliography}{119}
\providecommand{\natexlab}[1]{#1}
\providecommand{\url}[1]{\texttt{#1}}
\expandafter\ifx\csname urlstyle\endcsname\relax
  \providecommand{\doi}[1]{doi: #1}\else
  \providecommand{\doi}{doi: \begingroup \urlstyle{rm}\Url}\fi

\bibitem[Adamic and Adar(2003)]{adamic2003friends}
L.~A. Adamic and E.~Adar.
\newblock Friends and neighbors on the web.
\newblock \emph{Social networks}, 25\penalty0 (3):\penalty0 211--230, 2003.

\bibitem[Aggarwal(2011)]{Aggarwal2011}
C.~C. Aggarwal.
\newblock {An Introduction to Social Network Data Analytics}.
\newblock In C.~C. Aggarwal, editor, \emph{Social Network Data Analytics},
  chapter~1, pages 1--15. SPRINGER, 2011.
\newblock \doi{10.1007/978-1-4419-8462-3_1}.

\bibitem[Altman(1992)]{altman1992introduction}
N.~S. Altman.
\newblock An introduction to kernel and nearest-neighbor nonparametric
  regression.
\newblock \emph{The American Statistician}, 46\penalty0 (3):\penalty0 175--185,
  1992.

\bibitem[Amato et~al.(2016)Amato, Cozzolino, {Di Martino}, Mazzeo, Moscato,
  Picariello, Romano, and Sperl{\'{i}}]{Amato2016577}
F.~Amato, G.~Cozzolino, S.~{Di Martino}, A.~Mazzeo, V.~Moscato, A.~Picariello,
  S.~Romano, and G.~Sperl{\'{i}}.
\newblock {Opinions analysis in social networks for cultural heritage
  applications}.
\newblock \emph{Smart Innovation, Systems and Technologies}, 55:\penalty0
  577--586, 2016.
\newblock ISSN 21903018.
\newblock \doi{10.1007/978-3-319-39345-2_51}.

\bibitem[Arlot and Celisse(2010)]{arlot2010survey}
S.~Arlot and A.~Celisse.
\newblock A survey of cross-validation procedures for model selection.
\newblock \emph{Statistics surveys}, 4:\penalty0 40--79, 2010.

\bibitem[Bai et~al.(2021{\natexlab{a}})Bai, Luo, Nourian, and
  Pereira~Roders]{bai-etal-2021-whose-heritage}
N.~Bai, R.~Luo, P.~Nourian, and A.~Pereira~Roders.
\newblock {WHOS}e {H}eritage: {C}lassification of {UNESCO} {W}orld {H}eritage
  statements of ''{O}utstanding {U}niversal {V}alue{''} with soft labels.
\newblock In \emph{Findings of the Association for Computational Linguistics:
  EMNLP 2021}, pages 366--384, Punta Cana, Dominican Republic, Nov.
  2021{\natexlab{a}}. Association for Computational Linguistics.

\bibitem[Bai et~al.(2021{\natexlab{b}})Bai, Nourian, Luo, and
  Pereira~Roders]{Bai2021}
N.~Bai, P.~Nourian, R.~Luo, and A.~Pereira~Roders.
\newblock “{W}hat is {OUV}” revisited: A computational interpretation on
  the statements of {O}utstanding {U}niversal {V}alue.
\newblock \emph{ISPRS Annals of the Photogrammetry, Remote Sensing and Spatial
  Information Sciences}, VIII-M-1-2021:\penalty0 25--32, 2021{\natexlab{b}}.
\newblock \doi{10.5194/isprs-annals-VIII-M-1-2021-25-2021}.

\bibitem[Bai et~al.(2021{\natexlab{c}})Bai, Nourian, and
  Pereira~Roders]{bai2021_2}
N.~Bai, P.~Nourian, and A.~Pereira~Roders.
\newblock Global citizens and world heritage: Social inclusion of online
  communities in heritage planning.
\newblock \emph{The International Archives of the Photogrammetry, Remote
  Sensing and Spatial Information Sciences}, XLVI-M-1-2021:\penalty0 23--30,
  2021{\natexlab{c}}.
\newblock \doi{10.5194/isprs-archives-XLVI-M-1-2021-23-2021}.

\bibitem[Baltrusaitis et~al.(2019)Baltrusaitis, Ahuja, and
  Morency]{Baltrusaitis2019}
T.~Baltrusaitis, C.~Ahuja, and L.~P. Morency.
\newblock {Multimodal Machine Learning: A Survey and Taxonomy}.
\newblock \emph{IEEE Transactions on Pattern Analysis and Machine
  Intelligence}, 41\penalty0 (2):\penalty0 423--443, 2019.
\newblock ISSN 19393539.
\newblock \doi{10.1109/TPAMI.2018.2798607}.

\bibitem[Barab{\'a}si(2013)]{barabasi2013network}
A.-L. Barab{\'a}si.
\newblock Network science.
\newblock \emph{Philosophical Transactions of the Royal Society A:
  Mathematical, Physical and Engineering Sciences}, 371\penalty0
  (1987):\penalty0 20120375, 2013.

\bibitem[Batty(2013)]{batty2013new}
M.~Batty.
\newblock \emph{The new science of cities}.
\newblock MIT press, 2013.

\bibitem[Blanchard and Volchenkov(2008)]{blanchard2008mathematical}
P.~Blanchard and D.~Volchenkov.
\newblock \emph{Mathematical analysis of urban spatial networks}.
\newblock Springer Science \& Business Media, 2008.

\bibitem[Blum and Mitchell(1998)]{blum1998combining}
A.~Blum and T.~Mitchell.
\newblock Combining labeled and unlabeled data with co-training.
\newblock In \emph{Proceedings of the eleventh annual conference on
  Computational learning theory}, pages 92--100, 1998.

\bibitem[Boeing(2017)]{boeing2017osmnx}
G.~Boeing.
\newblock Osmnx: New methods for acquiring, constructing, analyzing, and
  visualizing complex street networks.
\newblock \emph{Computers, Environment and Urban Systems}, 65:\penalty0
  126--139, 2017.

\bibitem[Bonci et~al.(2018)Bonci, Clini, Martin, Pirani, Quattrini, and
  Raikov]{Bonci2018305}
A.~Bonci, P.~Clini, R.~Martin, M.~Pirani, R.~Quattrini, and A.~Raikov.
\newblock {Collaborative intelligence cyber-physical system for the
  valorization and re-use of cultural heritage}.
\newblock \emph{Journal of Information Technology in Construction}, 23\penalty0
  (1):\penalty0 305--323, 2018.
\newblock ISSN 18744753.

\bibitem[Breiman(1996{\natexlab{a}})]{breiman1996bagging}
L.~Breiman.
\newblock Bagging predictors.
\newblock \emph{Machine learning}, 24\penalty0 (2):\penalty0 123--140,
  1996{\natexlab{a}}.

\bibitem[Breiman(1996{\natexlab{b}})]{breiman1996stacked}
L.~Breiman.
\newblock Stacked regressions.
\newblock \emph{Machine learning}, 24\penalty0 (1):\penalty0 49--64,
  1996{\natexlab{b}}.

\bibitem[Breiman(2001)]{breiman2001random}
L.~Breiman.
\newblock Random forests.
\newblock \emph{Machine learning}, 45\penalty0 (1):\penalty0 5--32, 2001.

\bibitem[Campillo-Alhama and Martinez-Sala(2019)]{Campillo2019}
C.~Campillo-Alhama and A.-M. Martinez-Sala.
\newblock {Events 2.0 in the transmedia branding strategy of World Cultural
  Heritage Sites}.
\newblock \emph{PROFESIONAL DE LA INFORMACION}, 28\penalty0 (5), 2019.
\newblock ISSN 1386-6710.
\newblock \doi{10.3145/epi.2019.sep.09}.

\bibitem[Cao et~al.(2018)Cao, Shen, Xie, Parkhi, and
  Zisserman]{cao2018vggface2}
Q.~Cao, L.~Shen, W.~Xie, O.~M. Parkhi, and A.~Zisserman.
\newblock Vggface2: A dataset for recognising faces across pose and age.
\newblock In \emph{2018 13th IEEE international conference on automatic face \&
  gesture recognition (FG 2018)}, pages 67--74. IEEE, 2018.

\bibitem[Cheng and Wicks(2014)]{cheng2014event}
T.~Cheng and T.~Wicks.
\newblock Event detection using twitter: A spatio-temporal approach.
\newblock \emph{PloS one}, 9\penalty0 (6):\penalty0 e97807, 2014.

\bibitem[Cho et~al.(2022)Cho, Kang, Yoon, Park, and Kim]{cho2022classifying}
N.~Cho, Y.~Kang, J.~Yoon, S.~Park, and J.~Kim.
\newblock Classifying tourists’ photos and exploring tourism destination
  image using a deep learning model.
\newblock \emph{Journal of Quality Assurance in Hospitality \& Tourism}, pages
  1--29, 2022.

\bibitem[Chua et~al.(2009)Chua, Tang, Hong, Li, Luo, and Zheng]{Chua2009}
T.-S. Chua, J.~Tang, R.~Hong, H.~Li, Z.~Luo, and Y.~Zheng.
\newblock Nus-wide: a real-world web image database from national university of
  singapore.
\newblock In \emph{Proceedings of the ACM international conference on image and
  video retrieval}, pages 1--9, 2009.

\bibitem[Clark et~al.(2019)Clark, Khandelwal, Levy, and Manning]{Clark2019}
K.~Clark, U.~Khandelwal, O.~Levy, and C.~D. Manning.
\newblock {What does BERT look at? An analysis of BERT's attention}.
\newblock \emph{arXiv}, pages 276--286, 2019.
\newblock \doi{10.18653/v1/w19-4828}.

\bibitem[Crandall et~al.(2009)Crandall, Backstrom, Huttenlocher, and
  Kleinberg]{Crandall2009}
D.~Crandall, L.~Backstrom, D.~Huttenlocher, and J.~Kleinberg.
\newblock {Mapping the world's photos}.
\newblock \emph{WWW'09 - Proceedings of the 18th International World Wide Web
  Conference}, pages 761--770, 2009.
\newblock \doi{10.1145/1526709.1526812}.

\bibitem[Deng et~al.(2009)Deng, Dong, Socher, Li, Li, and
  Fei-Fei]{deng2009imagenet}
J.~Deng, W.~Dong, R.~Socher, L.-J. Li, K.~Li, and L.~Fei-Fei.
\newblock Imagenet: A large-scale hierarchical image database.
\newblock In \emph{2009 IEEE conference on computer vision and pattern
  recognition}, pages 248--255. Ieee, 2009.

\bibitem[Devlin et~al.(2019)Devlin, Chang, Lee, and Toutanova]{devlin2019bert}
J.~Devlin, M.-W. Chang, K.~Lee, and K.~Toutanova.
\newblock Bert: Pre-training of deep bidirectional transformers for language
  understanding.
\newblock In \emph{Proceedings of the 2019 Conference of the North American
  Chapter of the Association for Computational Linguistics: Human Language
  Technologies, Volume 1 (Long and Short Papers)}, pages 4171--4186, 2019.

\bibitem[Eom and Jo(2015)]{eom2015tail}
Y.-H. Eom and H.-H. Jo.
\newblock Tail-scope: Using friends to estimate heavy tails of degree
  distributions in large-scale complex networks.
\newblock \emph{Scientific reports}, 5\penalty0 (1):\penalty0 1--9, 2015.

\bibitem[Gabrielli et~al.(2014)Gabrielli, Rinzivillo, Ronzano, and
  Villatoro]{Gabrielli2014}
L.~Gabrielli, S.~Rinzivillo, F.~Ronzano, and D.~Villatoro.
\newblock {From Tweets to Semantic Trajectories: Mining Anomalous Urban
  Mobility Patterns}.
\newblock In D.~{Nin, J and Villatoro}, editor, \emph{CITIZEN IN SENSOR
  NETWORKS}, volume 8313 of \emph{Lecture Notes in Artificial Intelligence},
  pages 26--35, HEIDELBERGER PLATZ 3, D-14197 BERLIN, GERMANY, 2014.
  SPRINGER-VERLAG BERLIN.
\newblock ISBN 978-3-319-04178-0; 978-3-319-04177-3.
\newblock \doi{10.1007/978-3-319-04178-0_3}.

\bibitem[Gao and Ji(2019)]{gao2019graph}
H.~Gao and S.~Ji.
\newblock Graph u-nets.
\newblock In \emph{international conference on machine learning}, pages
  2083--2092. PMLR, 2019.

\bibitem[Giglio et~al.(2019{\natexlab{a}})Giglio, Bertacchini, Bilotta, and
  Pantano]{Giglio2019}
S.~Giglio, F.~Bertacchini, E.~Bilotta, and P.~Pantano.
\newblock {Machine learning and point of interests: typical tourist Italian
  cities}.
\newblock \emph{Current Issues in Tourism}, 0\penalty0 (0):\penalty0 1--13,
  2019{\natexlab{a}}.
\newblock ISSN 13683500.
\newblock \doi{10.1080/13683500.2019.1637827}.

\bibitem[Giglio et~al.(2019{\natexlab{b}})Giglio, Bertacchini, Bilotta, and
  Pantano]{Giglio2019306}
S.~Giglio, F.~Bertacchini, E.~Bilotta, and P.~Pantano.
\newblock {Using social media to identify tourism attractiveness in six Italian
  cities}.
\newblock \emph{Tourism Management}, 72:\penalty0 306--312, 2019{\natexlab{b}}.
\newblock ISSN 02615177.
\newblock \doi{10.1016/j.tourman.2018.12.007}.

\bibitem[Ginzarly et~al.(2019)Ginzarly, {Pereira Roders}, and
  Teller]{Ginzarly2019}
M.~Ginzarly, A.~{Pereira Roders}, and J.~Teller.
\newblock {Mapping historic urban landscape values through social media}.
\newblock \emph{Journal of Cultural Heritage}, 36:\penalty0 1--11, 2019.
\newblock ISSN 12962074.
\newblock \doi{10.1016/j.culher.2018.10.002}.

\bibitem[Gomez et~al.(2019)Gomez, Gomez, Gibert, and Karatzas]{Gomez2019530}
R.~Gomez, L.~Gomez, J.~Gibert, and D.~Karatzas.
\newblock {Learning from \#barcelona instagram data what locals and tourists
  post about its neighbourhoods}.
\newblock \emph{Lecture Notes in Computer Science (including subseries Lecture
  Notes in Artificial Intelligence and Lecture Notes in Bioinformatics)}, 11134
  LNCS:\penalty0 530--544, 2019.
\newblock ISSN 03029743.
\newblock \doi{10.1007/978-3-030-11024-6_41}.

\bibitem[Goodfellow et~al.(2014)Goodfellow, Pouget-Abadie, Mirza, Xu,
  Warde-Farley, Ozair, Courville, and Bengio]{goodfellow2014generative}
I.~Goodfellow, J.~Pouget-Abadie, M.~Mirza, B.~Xu, D.~Warde-Farley, S.~Ozair,
  A.~Courville, and Y.~Bengio.
\newblock Generative adversarial nets.
\newblock \emph{Advances in neural information processing systems}, 27, 2014.

\bibitem[Gustcoven(2016)]{gustcoven2016attributes}
E.~Gustcoven.
\newblock Attributes of world heritage cities, sustainability by management--a
  comparative study between the world heritage cities of amsterdam, edinburgh
  and quer{\'e}taro.
\newblock Master's thesis, KU Leuven, 2016.

\bibitem[Hagberg et~al.(2008)Hagberg, Swart, and S~Chult]{hagberg2008exploring}
A.~Hagberg, P.~Swart, and D.~S~Chult.
\newblock Exploring network structure, dynamics, and function using networkx.
\newblock Technical report, Los Alamos National Lab.(LANL), Los Alamos, NM
  (United States), 2008.

\bibitem[He et~al.(2016)He, Zhang, Ren, and Sun]{he2016deep}
K.~He, X.~Zhang, S.~Ren, and J.~Sun.
\newblock Deep residual learning for image recognition.
\newblock In \emph{Proceedings of the IEEE conference on computer vision and
  pattern recognition}, pages 770--778, 2016.

\bibitem[Hillier and Hanson(1989)]{hillier1989social}
B.~Hillier and J.~Hanson.
\newblock \emph{The Social Logic of Space}.
\newblock Cambridge University Press, 1989.
\newblock ISBN 9781139935685.

\bibitem[Hinton(1990)]{hinton1990connectionist}
G.~E. Hinton.
\newblock Connectionist learning procedures.
\newblock In \emph{Machine learning}, pages 555--610. Elsevier, 1990.

\bibitem[Howard and Ruder(2018)]{ulmfit}
J.~Howard and S.~Ruder.
\newblock Universal language model fine-tuning for text classification.
\newblock In I.~Gurevych and Y.~Miyao, editors, \emph{Proceedings of the 56th
  Annual Meeting of the Association for Computational Linguistics, {ACL} 2018,
  Melbourne, Australia, July 15-20, 2018, Volume 1: Long Papers}, pages
  328--339. Association for Computational Linguistics, 2018.
\newblock \doi{10.18653/v1/P18-1031}.

\bibitem[Howley et~al.(2009)Howley, Scott, and
  Redmond]{howley2009sustainability}
P.~Howley, M.~Scott, and D.~Redmond.
\newblock Sustainability versus liveability: an investigation of neighbourhood
  satisfaction.
\newblock \emph{Journal of environmental planning and management}, 52\penalty0
  (6):\penalty0 847--864, 2009.

\bibitem[Huiskes and Lew(2008)]{Huiskes2008}
M.~J. Huiskes and M.~S. Lew.
\newblock {The MIR Flickr retrieval evaluation}.
\newblock \emph{Proceedings of the 1st International ACM Conference on
  Multimedia Information Retrieval, MIR2008, Co-located with the 2008 ACM
  International Conference on Multimedia, MM'08}, pages 39--43, 2008.
\newblock \doi{10.1145/1460096.1460104}.

\bibitem[Jokilehto(2007)]{jokilehto2007ouv}
J.~Jokilehto.
\newblock Aesthetics in the world heritage context.
\newblock In \emph{Values and Criteria in Heritage Conservation}, pages
  183--194. Polistampa, 2007.

\bibitem[Jokilehto(2008)]{Jokilehto2008}
J.~Jokilehto.
\newblock {What is OUV? Defining the Outstanding Universal Value of Cultural
  World Heritage Properties}.
\newblock Technical report, ICOMOS, ICOMOS Berlin, 2008.

\bibitem[Kang et~al.(2021)Kang, Cho, Yoon, Park, and Kim]{kang2021transfer}
Y.~Kang, N.~Cho, J.~Yoon, S.~Park, and J.~Kim.
\newblock Transfer learning of a deep learning model for exploring tourists’
  urban image using geotagged photos.
\newblock \emph{ISPRS International Journal of Geo-Information}, 10\penalty0
  (3):\penalty0 137, 2021.

\bibitem[Karypis and Kumar(1995)]{karypis1995analysis}
G.~Karypis and V.~Kumar.
\newblock Analysis of multilevel graph partitioning.
\newblock In \emph{Supercomputing'95: Proceedings of the 1995 ACM/IEEE
  conference on Supercomputing}, pages 29--29. IEEE, 1995.

\bibitem[Kipf and Welling(2016)]{kipf2016semi}
T.~N. Kipf and M.~Welling.
\newblock Semi-supervised classification with graph convolutional networks.
\newblock \emph{arXiv preprint arXiv:1609.02907}, 2016.

\bibitem[Krizhevsky et~al.(2012)Krizhevsky, Sutskever, and
  Hinton]{krizhevsky2012imagenet}
A.~Krizhevsky, I.~Sutskever, and G.~E. Hinton.
\newblock Imagenet classification with deep convolutional neural networks.
\newblock \emph{Advances in neural information processing systems}, 25, 2012.

\bibitem[Lafon and Lee(2006)]{lafon2006diffusion}
S.~Lafon and A.~B. Lee.
\newblock Diffusion maps and coarse-graining: A unified framework for
  dimensionality reduction, graph partitioning, and data set parameterization.
\newblock \emph{IEEE transactions on pattern analysis and machine
  intelligence}, 28\penalty0 (9):\penalty0 1393--1403, 2006.

\bibitem[Lazer et~al.(2009)Lazer, Pentland, Adamic, Aral, Barabasi, Brewer,
  Christakis, Contractor, Fowler, Gutmann, et~al.]{lazer2009social}
D.~Lazer, A.~Pentland, L.~Adamic, S.~Aral, A.-L. Barabasi, D.~Brewer,
  N.~Christakis, N.~Contractor, J.~Fowler, M.~Gutmann, et~al.
\newblock Social science. computational social science.
\newblock \emph{Science (New York, NY)}, 323\penalty0 (5915):\penalty0
  721--723, 2009.

\bibitem[Lee et~al.(2013)]{lee2013pseudo}
D.-H. Lee et~al.
\newblock Pseudo-label: The simple and efficient semi-supervised learning
  method for deep neural networks.
\newblock In \emph{Workshop on challenges in representation learning, ICML},
  volume~3, page 896, 2013.

\bibitem[Lee and Kang(2021)]{lee2021mining}
H.~Lee and Y.~Kang.
\newblock Mining tourists’ destinations and preferences through lstm-based
  text classification and spatial clustering using flickr data.
\newblock \emph{Spatial Information Research}, 29\penalty0 (6):\penalty0
  825--839, 2021.

\bibitem[Liew(2014)]{Liew2014}
C.~L. Liew.
\newblock {Participatory cultural heritage: A tale of two institutions' use of
  social media}.
\newblock \emph{D-Lib Magazine}, 20\penalty0 (3-4):\penalty0 1--17, 2014.
\newblock ISSN 10829873.
\newblock \doi{10.1045/march2014-liew}.

\bibitem[Lin et~al.(2014)Lin, Maire, Belongie, Hays, Perona, Ramanan,
  Doll{\'a}r, and Zitnick]{lin2014microsoft}
T.-Y. Lin, M.~Maire, S.~Belongie, J.~Hays, P.~Perona, D.~Ramanan,
  P.~Doll{\'a}r, and C.~L. Zitnick.
\newblock Microsoft coco: Common objects in context.
\newblock In \emph{European conference on computer vision}, pages 740--755.
  Springer, 2014.

\bibitem[Lu and Stepchenkova(2015)]{Lu2015}
W.~Lu and S.~Stepchenkova.
\newblock {User-Generated Content as a Research Mode in Tourism and Hospitality
  Applications: Topics, Methods, and Software}.
\newblock \emph{Journal of Hospitality Marketing and Management}, 24\penalty0
  (2):\penalty0 119--154, 2015.
\newblock ISSN 19368631.
\newblock \doi{10.1080/19368623.2014.907758}.

\bibitem[Ma and Tang(2021)]{ma2021deep}
Y.~Ma and J.~Tang.
\newblock \emph{Deep learning on graphs}.
\newblock Cambridge University Press, 2021.

\bibitem[Ma et~al.(2018)Ma, Ren, Jiang, Tang, and Yin]{ma2018multi}
Y.~Ma, Z.~Ren, Z.~Jiang, J.~Tang, and D.~Yin.
\newblock Multi-dimensional network embedding with hierarchical structure.
\newblock In \emph{Proceedings of the eleventh ACM international conference on
  web search and data mining}, pages 387--395, 2018.

\bibitem[Majid et~al.(2013)Majid, Chen, Chen, Mirza, Hussain, and
  Woodward]{majid2013context}
A.~Majid, L.~Chen, G.~Chen, H.~T. Mirza, I.~Hussain, and J.~Woodward.
\newblock A context-aware personalized travel recommendation system based on
  geotagged social media data mining.
\newblock \emph{International Journal of Geographical Information Science},
  27\penalty0 (4):\penalty0 662--684, 2013.

\bibitem[Marine-Roig and {Anton Clav{\'{e}}}(2015)]{Marine-Roig2015}
E.~Marine-Roig and S.~{Anton Clav{\'{e}}}.
\newblock {Tourism analytics with massive user-generated content: A case study
  of Barcelona}.
\newblock \emph{Journal of Destination Marketing and Management}, 4\penalty0
  (3):\penalty0 162--172, 2015.
\newblock ISSN 2212571X.
\newblock \doi{10.1016/j.jdmm.2015.06.004}.

\bibitem[Miah et~al.(2017)Miah, Vu, Gammack, and McGrath]{miah2017big}
S.~J. Miah, H.~Q. Vu, J.~Gammack, and M.~McGrath.
\newblock A big data analytics method for tourist behaviour analysis.
\newblock \emph{Information \& Management}, 54\penalty0 (6):\penalty0 771--785,
  2017.

\bibitem[Monteiro et~al.(2014)Monteiro, Henriques, Painho, and
  Vaz]{Monteiro2014}
V.~Monteiro, R.~Henriques, M.~Painho, and E.~Vaz.
\newblock {Sensing World Heritage An Exploratory Study of Twitter as a Tool for
  Assessing Reputation}.
\newblock In O.~{Murgante, B and Misra, S and Rocha, AMAC and Torre, C and
  Rocha, JG and Falcao, MI and Taniar, D and Apduhan, BO and Gervasi}, editor,
  \emph{COMPUTATIONAL SCIENCE AND ITS APPLICATIONS - ICCSA 2014, PT II}, volume
  8580 of \emph{Lecture Notes in Computer Science}, pages 404--419. Univ Minho;
  Univ Perugia; Univ Basilicata; Monash Univ; Kyushu Sangyo Univ; Assoc
  Portuguesa Investigacao Operac, 2014.
\newblock ISBN 978-3-319-09129-7; 978-3-319-09128-0.

\bibitem[Nguyen et~al.(2018)Nguyen, Lee, Rossi, Ahmed, Koh, and
  Kim]{nguyen2018continuous}
G.~H. Nguyen, J.~B. Lee, R.~A. Rossi, N.~K. Ahmed, E.~Koh, and S.~Kim.
\newblock Continuous-time dynamic network embeddings.
\newblock In \emph{Companion Proceedings of the The Web Conference 2018}, pages
  969--976, 2018.

\bibitem[Nourian(2016)]{nourian2016configraphics}
P.~Nourian.
\newblock \emph{Configraphics: Graph Theoretical Methods for Design and
  Analysis of Spatial Configurations}.
\newblock TU Delft, 2016.
\newblock ISBN 9789461867209.

\bibitem[Nourian et~al.(2016)Nourian, Rezvani, Sariyildiz, and van~der
  Hoeven]{nourian2016spectral}
P.~Nourian, S.~Rezvani, I.~Sariyildiz, and F.~van~der Hoeven.
\newblock Spectral modelling for spatial network analysis.
\newblock In \emph{Proceedings of the Symposium on Simulation for Architecture
  and Urban Design (simAUD 2016)}, pages 103--110. SimAUD, 2016.

\bibitem[Nowak and R{\"{u}}ger(2010)]{Nowak2010}
S.~Nowak and S.~R{\"{u}}ger.
\newblock {How reliable are annotations via crowdsourcing}.
\newblock In \emph{Proceedings of the international conference on Multimedia
  information retrieval}, pages 557--566, 2010.
\newblock ISBN 9781605588155.
\newblock \doi{10.1145/1743384.1743478}.

\bibitem[Pan and Yang(2009)]{pan2009survey}
S.~J. Pan and Q.~Yang.
\newblock A survey on transfer learning.
\newblock \emph{IEEE Transactions on knowledge and data engineering},
  22\penalty0 (10):\penalty0 1345--1359, 2009.

\bibitem[Pang et~al.(2021)Pang, Zhao, and Li]{pang2021graph}
Y.~Pang, Y.~Zhao, and D.~Li.
\newblock Graph pooling via coarsened graph infomax.
\newblock In \emph{Proceedings of the 44th International ACM SIGIR Conference
  on Research and Development in Information Retrieval}, pages 2177--2181,
  2021.

\bibitem[Patterson and Hays(2012)]{patterson2012sun}
G.~Patterson and J.~Hays.
\newblock Sun attribute database: Discovering, annotating, and recognizing
  scene attributes.
\newblock In \emph{2012 IEEE Conference on Computer Vision and Pattern
  Recognition}, pages 2751--2758. IEEE, 2012.

\bibitem[Patterson et~al.(2014)Patterson, Xu, Su, and Hays]{patterson2014sun}
G.~Patterson, C.~Xu, H.~Su, and J.~Hays.
\newblock The sun attribute database: Beyond categories for deeper scene
  understanding.
\newblock \emph{International Journal of Computer Vision}, 108\penalty0
  (1):\penalty0 59--81, 2014.

\bibitem[Pearson(1905)]{pearson1905problem}
K.~Pearson.
\newblock The problem of the random walk.
\newblock \emph{Nature}, 72\penalty0 (1865):\penalty0 294--294, 1905.

\bibitem[Pedregosa et~al.(2011)Pedregosa, Varoquaux, Gramfort, Michel, Thirion,
  Grisel, Blondel, Prettenhofer, Weiss, Dubourg, Vanderplas, Passos,
  Cournapeau, Brucher, Perrot, and Duchesnay]{scikit-learn}
F.~Pedregosa, G.~Varoquaux, A.~Gramfort, V.~Michel, B.~Thirion, O.~Grisel,
  M.~Blondel, P.~Prettenhofer, R.~Weiss, V.~Dubourg, J.~Vanderplas, A.~Passos,
  D.~Cournapeau, M.~Brucher, M.~Perrot, and E.~Duchesnay.
\newblock Scikit-learn: Machine learning in {P}ython.
\newblock \emph{Journal of Machine Learning Research}, 12:\penalty0 2825--2830,
  2011.

\bibitem[Penn(2003)]{penn2003space}
A.~Penn.
\newblock Space syntax and spatial cognition: or why the axial line?
\newblock \emph{Environment and behavior}, 35\penalty0 (1):\penalty0 30--65,
  2003.

\bibitem[Pentland(2015)]{pentland2015social}
A.~Pentland.
\newblock \emph{Social Physics: How social networks can make us smarter}.
\newblock Penguin, 2015.

\bibitem[{Pereira Roders}(2007)]{PereiraRoders2007}
A.~{Pereira Roders}.
\newblock \emph{{Re-architecture: lifespan rehabilitation of built heritage.}}
\newblock PhD thesis, Technische Universiteit Eindhoven, 2007.

\bibitem[{Pereira Roders}(2010)]{PereiraRoders2010}
A.~{Pereira Roders}.
\newblock {Revealing the World Heritage cities and their varied natures}.
\newblock In \emph{Heritage 2010: Heritage and Sustainable Development, Vols 1
  and 2}, chapter Heritage a, pages 245--253. Green Lines Institute, 2010.
\newblock ISBN 978-989-95671-3-9.

\bibitem[{Pereira Roders}(2019)]{PereiraRoders2019}
A.~{Pereira Roders}.
\newblock {The Historic Urban Landscape Approach in Action: Eight Years Later}.
\newblock In \emph{Reshaping Urban Conservation}, pages 21--54. Springer, 2019.
\newblock ISBN 9789811088872.
\newblock \doi{10.1007/978-981-10-8887-2_2}.

\bibitem[Pickering et~al.(2018)Pickering, Rossi, Hernando, and
  Barros]{Pickering2018}
C.~Pickering, S.~D. Rossi, A.~Hernando, and A.~Barros.
\newblock {Current knowledge and future research directions for the monitoring
  and management of visitors in recreational and protected areas}.
\newblock \emph{Journal of Outdoor Recreation and Tourism}, 21\penalty0
  (November 2017):\penalty0 10--18, 2018.
\newblock ISSN 22130780.
\newblock \doi{10.1016/j.jort.2017.11.002}.

\bibitem[Platt et~al.(1999)]{platt1999probabilistic}
J.~Platt et~al.
\newblock Probabilistic outputs for support vector machines and comparisons to
  regularized likelihood methods.
\newblock \emph{Advances in large margin classifiers}, 10\penalty0
  (3):\penalty0 61--74, 1999.

\bibitem[Plummer et~al.(2015)Plummer, Wang, Cervantes, Caicedo, Hockenmaier,
  and Lazebnik]{plummer2015flickr30k}
B.~A. Plummer, L.~Wang, C.~M. Cervantes, J.~C. Caicedo, J.~Hockenmaier, and
  S.~Lazebnik.
\newblock Flickr30k entities: Collecting region-to-phrase correspondences for
  richer image-to-sentence models.
\newblock In \emph{Proceedings of the IEEE international conference on computer
  vision}, pages 2641--2649, 2015.

\bibitem[Prince(2004)]{prince2004does}
M.~Prince.
\newblock Does active learning work? a review of the research.
\newblock \emph{Journal of engineering education}, 93\penalty0 (3):\penalty0
  223--231, 2004.

\bibitem[Pustejovsky and Stubbs(2012)]{pustejovsky2012natural}
J.~Pustejovsky and A.~Stubbs.
\newblock \emph{Natural Language Annotation for Machine Learning: A guide to
  corpus-building for applications}.
\newblock " O'Reilly Media, Inc.", 2012.

\bibitem[Rakic and Chambers(2008)]{Rakic2008}
T.~Rakic and D.~Chambers.
\newblock {World Heritage: Exploring the Tension Between the National and the
  ‘Universal'}.
\newblock \emph{Journal of Heritage Tourism}, 2\penalty0 (3):\penalty0
  145--155, 2008.
\newblock ISSN 17476631.
\newblock \doi{10.2167/jht056.0}.

\bibitem[Ratti(2004)]{Ratti2004}
C.~Ratti.
\newblock Space syntax: Some inconsistencies.
\newblock \emph{Environment and Planning B: Planning and Design}, 31\penalty0
  (4):\penalty0 487--499, 2004.
\newblock \doi{10.1068/b3019}.

\bibitem[Reiter(1989)]{reiter1989towards}
R.~Reiter.
\newblock Towards a logical reconstruction of relational database theory.
\newblock In \emph{Readings in Artificial Intelligence and Databases}, pages
  301--327. Elsevier, 1989.

\bibitem[Ren et~al.(2019)Ren, Cheng, and Zhang]{ren2019deep}
Y.~Ren, T.~Cheng, and Y.~Zhang.
\newblock Deep spatio-temporal residual neural networks for road-network-based
  data modeling.
\newblock \emph{International Journal of Geographical Information Science},
  33\penalty0 (9):\penalty0 1894--1912, 2019.

\bibitem[Rish et~al.(2001)]{rish2001empirical}
I.~Rish et~al.
\newblock An empirical study of the naive bayes classifier.
\newblock In \emph{IJCAI 2001 workshop on empirical methods in artificial
  intelligence}, volume~3, pages 41--46, 2001.

\bibitem[Schroff et~al.(2015)Schroff, Kalenichenko, and
  Philbin]{schroff2015facenet}
F.~Schroff, D.~Kalenichenko, and J.~Philbin.
\newblock Facenet: A unified embedding for face recognition and clustering.
\newblock In \emph{Proceedings of the IEEE conference on computer vision and
  pattern recognition}, pages 815--823, 2015.

\bibitem[Settles(2011)]{settles2011theories}
B.~Settles.
\newblock From theories to queries: Active learning in practice.
\newblock In \emph{Active learning and experimental design workshop in
  conjunction with AISTATS 2010}, pages 1--18. JMLR Workshop and Conference
  Proceedings, 2011.

\bibitem[Sohn et~al.(2020)Sohn, Berthelot, Carlini, Zhang, Zhang, Raffel,
  Cubuk, Kurakin, and Li]{sohn2020fixmatch}
K.~Sohn, D.~Berthelot, N.~Carlini, Z.~Zhang, H.~Zhang, C.~A. Raffel, E.~D.
  Cubuk, A.~Kurakin, and C.-L. Li.
\newblock Fixmatch: Simplifying semi-supervised learning with consistency and
  confidence.
\newblock \emph{Advances in Neural Information Processing Systems},
  33:\penalty0 596--608, 2020.

\bibitem[Sun et~al.(2019)Sun, Qiu, Xu, and Huang]{Sun2019}
C.~Sun, X.~Qiu, Y.~Xu, and X.~Huang.
\newblock {How to Fine-Tune BERT for Text Classification?}
\newblock \emph{Lecture Notes in Computer Science (including subseries Lecture
  Notes in Artificial Intelligence and Lecture Notes in Bioinformatics)}, 11856
  LNAI\penalty0 (2):\penalty0 194--206, 2019.
\newblock ISSN 16113349.
\newblock \doi{10.1007/978-3-030-32381-3_16}.

\bibitem[Tang and Liu(2009)]{tang2009relational}
L.~Tang and H.~Liu.
\newblock Relational learning via latent social dimensions.
\newblock In \emph{Proceedings of the 15th ACM SIGKDD international conference
  on Knowledge discovery and data mining}, pages 817--826, 2009.

\bibitem[{Tarrafa Silva} and {Pereira Roders}(2010)]{TarrafaSilva2010}
A.~{Tarrafa Silva} and A.~{Pereira Roders}.
\newblock {The cultural significance of World Heritage cities : Portugal as
  case study}.
\newblock In \emph{Heritage and Sustainable Development}, pages 255--263,
  {\'{E}}vora, Portugal, 2010.
\newblock \doi{10.13140/2.1.1152.0800}.

\bibitem[Tobler(1970)]{tobler1970computer}
W.~R. Tobler.
\newblock A computer movie simulating urban growth in the detroit region.
\newblock \emph{Economic geography}, 46\penalty0 (sup1):\penalty0 234--240,
  1970.

\bibitem[UNESCO(1972)]{UNESCO1972}
UNESCO.
\newblock {Convention Concerning the Protection of the World Cultural and
  Natural Heritage}.
\newblock Technical Report november, UNESCO, Paris, 1972.

\bibitem[UNESCO(2008)]{UNESCO2008}
UNESCO.
\newblock {Operational guidelines for the implementation of the world heritage
  convention}.
\newblock Technical Report July, UNESCO World Heritage Centre, 2008.

\bibitem[UNESCO(2011)]{UNESCO2011}
UNESCO.
\newblock {RECOMMENDATION ON THE HISTORIC URBAN LANDSCAPE}.
\newblock Technical report, UNESCO, Paris, 2011.

\bibitem[UNESCO(2020)]{UNESCO2020}
UNESCO.
\newblock {Heritage in Urban Contexts: Impact of Development Projects on World
  Heritage properties in Cities}.
\newblock Technical Report January, UNESCO World Heritage Centre, 2020.

\bibitem[Urry and Larsen(2011)]{urry2011tourist}
J.~Urry and J.~Larsen.
\newblock \emph{The tourist gaze 3.0}.
\newblock Sage, 2011.

\bibitem[Valese et~al.(2020)Valese, Noardo, and Pereira~Roders]{Valese2020}
M.~Valese, F.~Noardo, and A.~Pereira~Roders.
\newblock World heritage mapping in a standard-based structured geographical
  information system.
\newblock \emph{The International Archives of the Photogrammetry, Remote
  Sensing and Spatial Information Sciences}, XLIII-B4-2020:\penalty0 81--88,
  2020.
\newblock \doi{10.5194/isprs-archives-XLIII-B4-2020-81-2020}.

\bibitem[Vaswani et~al.(2017)Vaswani, Shazeer, Parmar, Uszkoreit, Jones, Gomez,
  Kaiser, and Polosukhin]{vaswani2017attention}
A.~Vaswani, N.~Shazeer, N.~Parmar, J.~Uszkoreit, L.~Jones, A.~N. Gomez,
  {\L}.~Kaiser, and I.~Polosukhin.
\newblock Attention is all you need.
\newblock In \emph{Proceedings of the 31st International Conference on Neural
  Information Processing Systems}, pages 6000--6010, 2017.

\bibitem[Veldpaus(2015)]{Veldpaus2015}
L.~Veldpaus.
\newblock \emph{{Historic urban landscapes: framing the integration of urban
  and heritage planning in multilevel governance}}.
\newblock PhD thesis, Technische Universiteit Eindhoven, 2015.

\bibitem[Veldpaus and Roders(2014)]{veldpaus2014learning}
L.~Veldpaus and A.~P. Roders.
\newblock Learning from a legacy: Venice to valletta.
\newblock \emph{Change over time}, 4\penalty0 (2):\penalty0 244--263, 2014.

\bibitem[Veli{\v{c}}kovi{\'c} et~al.(2017)Veli{\v{c}}kovi{\'c}, Cucurull,
  Casanova, Romero, Lio, and Bengio]{velivckovic2017graph}
P.~Veli{\v{c}}kovi{\'c}, G.~Cucurull, A.~Casanova, A.~Romero, P.~Lio, and
  Y.~Bengio.
\newblock Graph attention networks.
\newblock \emph{arXiv preprint arXiv:1710.10903}, 2017.

\bibitem[Wang et~al.(2020)Wang, Yao, Kwok, and Ni]{wang2020generalizing}
Y.~Wang, Q.~Yao, J.~T. Kwok, and L.~M. Ni.
\newblock Generalizing from a few examples: A survey on few-shot learning.
\newblock \emph{ACM computing surveys (csur)}, 53\penalty0 (3):\penalty0 1--34,
  2020.

\bibitem[Wasserman et~al.(1994)Wasserman, Faust, et~al.]{wasserman1994social}
S.~Wasserman, K.~Faust, et~al.
\newblock \emph{Social network analysis: Methods and applications}.
\newblock Cambridge university press, 1994.

\bibitem[Williams et~al.(2017)Williams, Inversini, Ferdinand, and
  Buhalis]{Williams201787}
N.~L. Williams, A.~Inversini, N.~Ferdinand, and D.~Buhalis.
\newblock {Destination eWOM: A macro and meso network approach?}
\newblock \emph{Annals of Tourism Research}, 64:\penalty0 87--101, 2017.
\newblock ISSN 01607383.
\newblock \doi{10.1016/j.annals.2017.02.007}.

\bibitem[Yuster and Zwick(2005)]{yuster2005fast}
R.~Yuster and U.~Zwick.
\newblock Fast sparse matrix multiplication.
\newblock \emph{ACM Transactions On Algorithms (TALG)}, 1\penalty0
  (1):\penalty0 2--13, 2005.

\bibitem[Zeng et~al.(2019)Zeng, Zhou, Srivastava, Kannan, and
  Prasanna]{zeng2019graphsaint}
H.~Zeng, H.~Zhou, A.~Srivastava, R.~Kannan, and V.~Prasanna.
\newblock Graphsaint: Graph sampling based inductive learning method.
\newblock \emph{arXiv preprint arXiv:1907.04931}, 2019.

\bibitem[Zhang et~al.(2019)Zhang, Zhou, Ratti, and Liu]{zhang2019discovering}
F.~Zhang, B.~Zhou, C.~Ratti, and Y.~Liu.
\newblock Discovering place-informative scenes and objects using social media
  photos.
\newblock \emph{Royal Society open science}, 6\penalty0 (3):\penalty0 181375,
  2019.

\bibitem[Zhang et~al.(2018)Zhang, Cui, Neumann, and Chen]{zhang2018end}
M.~Zhang, Z.~Cui, M.~Neumann, and Y.~Chen.
\newblock An end-to-end deep learning architecture for graph classification.
\newblock In \emph{Thirty-second AAAI conference on artificial intelligence},
  pages 4438--4445, 2018.

\bibitem[Zhang and Cheng(2020)]{zhang2020graph}
Y.~Zhang and T.~Cheng.
\newblock Graph deep learning model for network-based predictive hotspot
  mapping of sparse spatio-temporal events.
\newblock \emph{Computers, Environment and Urban Systems}, 79:\penalty0 101403,
  2020.

\bibitem[Zhang et~al.(2020)Zhang, Cui, and Zhu]{zhang2020deep}
Z.~Zhang, P.~Cui, and W.~Zhu.
\newblock Deep learning on graphs: A survey.
\newblock \emph{IEEE Transactions on Knowledge and Data Engineering}, 2020.

\bibitem[Zhou et~al.(2014)Zhou, Lapedriza, Xiao, Torralba, and
  Oliva]{zhou2014learning}
B.~Zhou, A.~Lapedriza, J.~Xiao, A.~Torralba, and A.~Oliva.
\newblock Learning deep features for scene recognition using places database.
\newblock \emph{Advances in neural information processing systems}, 27, 2014.

\bibitem[Zhou et~al.(2017)Zhou, Lapedriza, Khosla, Oliva, and
  Torralba]{zhou2017places}
B.~Zhou, A.~Lapedriza, A.~Khosla, A.~Oliva, and A.~Torralba.
\newblock Places: A 10 million image database for scene recognition.
\newblock \emph{IEEE transactions on pattern analysis and machine
  intelligence}, 40\penalty0 (6):\penalty0 1452--1464, 2017.

\bibitem[Zhou and Long(2016)]{zhou2016sinogrids}
Y.~Zhou and Y.~Long.
\newblock Sinogrids: a practice for open urban data in china.
\newblock \emph{Cartography and Geographic Information Science}, 43\penalty0
  (5):\penalty0 379--392, 2016.

\bibitem[Zhou(2012)]{zhou2012ensemble}
Z.-H. Zhou.
\newblock \emph{Ensemble methods: foundations and algorithms}.
\newblock CRC press, 2012.

\bibitem[Zhou and Li(2010)]{zhou2010semi}
Z.-H. Zhou and M.~Li.
\newblock Semi-supervised learning by disagreement.
\newblock \emph{Knowledge and Information Systems}, 24\penalty0 (3):\penalty0
  415--439, 2010.

\bibitem[Zhu and Goldberg(2009)]{zhu2009introduction}
X.~Zhu and A.~B. Goldberg.
\newblock Introduction to semi-supervised learning.
\newblock \emph{Synthesis lectures on artificial intelligence and machine
  learning}, 3\penalty0 (1):\penalty0 1--130, 2009.

\end{thebibliography}

\twocolumn
\appendix
\section*{Appendix}
\section{Abbreviations}
The following abbreviations are used in this manuscript:

\begin{tabular}{@{}lp{180px}}
Acc & Accuracy\\
AMS & Data of Amsterdam, the Netherlands\\
API & Application Programming Interface\\
BC-SVM & Bagging Classifier with SVMs as the internal base estimators\\
BERT & Bidirectional Encoder Representations from Transformers\\
CNN & Convolutional Neural Network\\
CC & Connected Components\\
CV & Cross-Validation\\
GNB & Gaussian Naive Bayes\\
HA & Heritage Attributes\\
HUL & Historic Urban Landscape\\
HV & Heritage Values\\
KNN & K-Nearest Neighbour\\
ML & Machine Learning\\
MLP & Multi-layer Perceptron\\
MML & Multi-modal Machine Learning\\
NLP & Natural Language Processing\\
OUV & Outstanding Universal Value\\
RF & Random Forest\\
SNA & Social Network Analysis\\
SUZ & Data of Suzhou, China\\
SVM & Support Vector Machine\\
UGC & User-Generated Content\\
ULMFiT & Universal Language Model Fine-tuning\\
UNESCO & The United Nations Educational, Scientific and Cultural Organization\\
VEN & Data of Venice, Italy\\
VEN-XL & The extra-large version of Venice data\\
w. & with\\
WH & World Heritage\\
WHL & World Heritage List\\
\end{tabular}

\section{Details of Collecting the Raw Dataset}\label{sec:App_data_collection}

For each case study city, FlickrAPI python library was used to access the \texttt{\small photo.search} API method provided by Flickr, using the Geo-locations in Table~\ref{T_Case_studies} as the centroids to search the IDs of geo-tagged images in a fixed radius (5km for Venice and Suzhou, and 2km for Amsterdam) that covers the major urban area of the corresponding UNESCO World Heritage property.
To make the datasets from the three cities comparable and compatible, a maximum of 5000 IDs was given to the search engine for each city, since Flickr users are relatively scarce in China.
For all the image IDs, only those with a flag of \texttt{\small candownload} indicated by the owner were further queried, in order to respect the privacy and copyrights of Flickr users.
Those images are further queried through \texttt{\small photo.getInfo} and \texttt{\small photo.getSizes} API methods to retrieve the following information: the owner's ID; the owner's registered location on Flickr; the title, description, and tags provided by the user; the geo-tag of the image; the timestamp of the image marking when it has been taken, and URLs to download the \texttt{\small Large Square} (150$\times$150 px) and \texttt{\small Small 320} (320$\times$240 px) versions of the original image.
The images that has the user tag of ``\emph{erotic}'' were excluded from the query.
Then all the images of both sizes were saved and transformed into RGB format as raw visual data.

A stop-word list has been used to remove the HTML symbols and other formatting elements from the texts and to filter out textual data that were mainly 1) a description of the camera used, 2) a default image name generated by the camera, 3) an advertisement or a promotion.
Google Translator API from the Deep Translator python library has been used to detect the languages of the posts on the sentence level to mark whether the sentence was written in English, the local language (Dutch, Chinese, and Italian respectively for the three cities), or any other languages.
The same API was used to translate the none-English sentences into English.
Then all the \emph{valid} sentences coming from any textual field of the same post were merged into a new field named \texttt{\small Revised Text} as the raw textual data.

Furthermore, the public contact (friend) lists and group (subscription) lists of all the retrieved owners have been queried through the \texttt{\small people.getPublicGroups} and the \texttt{\small contacts.getPublicList} API methods, while all user and group information were only considered as a [semi-] anonymous ID as respect to the privacy policy. 

To test the scalability of the methodological workflow, another larger dataset without the limit of maximum 5000 IDs has also been collected in Venice (VEN-XL).
The API of Flickr has a limitation at the scale of queries, which would return occasional errors while the server gets in burden.
This requires a different strategy during data collection of the larger dataset.
In this study, a $20\times 20$ grid was tiled in the area of Venice (from 45.420855N 12.291054E to 45.448286N 12.369234E) to collect the post IDs from the centroid of each tile with a radius of 0.3km, which were later aggregated by removing the duplicated IDs collected by multiple tiles to form the entire large dataset.
The further steps of data cleaning and pre-processing remained the same with the smaller datasets.

\section{Details for Machine Learning Models}\label{sec:App_ml}

A dataset with 902 sample images collected in Tripoli, Lebanon and classified with expert-based annotations presented in \cite{Ginzarly2019} was used to train several state-of-the-art ML models to replicate the experts' behaviour on classifying depicted scenery.
For each image, a unique class label among the 9 depicted scenes mentioned in Table~\ref{T_attr_definition} was provided.
10\% of the images were separated and kept away during training as test dataset and the remaining 812 images were used to train ML models with Scikit-learn python library \citep{scikit-learn}.
Among the 812 data points, \texttt{\small train\_test\_split} method was further used to split out a validation dataset with 203 samples (25\%).
The 512-dimensional visual representation introduced in Section~\ref{sec:vis feature} was generated from the images as the input of ML models, while the class label was used as an categorical output as multi-class single-label classification task.

For each of the selected ML models, \texttt{\small GridSearchCV} function with 10-fold validation was used to wrap the model with a set of tunable parameters in a small range to be selected, while the average top-1 accuracy was used as the criterion for model selection.
All 812 samples were input to the CV to tune the hyper-parameters, after which the trained models with their optimal hyper-parameters were tested on the 203 validation data samples and the unseen test dataset with the remained 90 samples.
For the latter steps, top-1 accuracy and macro-average F1 score (harmonic average of the precision and recall scores) of all classes were used as the evaluation metrics.
All experiments were conducted on 12th Gen Intel(R) Core(TM) i7-12700KF CPU.
The implementation details of the models are as follows:

\paragraph{MLP} The model used L2 penalty alpha of 1e-4, solver of stochastic gradient descent, adaptive learning rate and early stopping with maximum 300 iterations. It was tuned on the initial learning rate in $\{.05, .1, .2\}$, and the hidden layer sizes of one layer in $\{(32,),(64,),(128,),(256,),\}$ or two layers in $\{(256,128),(256,64),(256,32),(128,64),(128,32)\}$.
The best model had two hidden layers of (256,128) with learning rate of .05.

\paragraph{KNN} The model was tuned on the number of neighbours in range $[3,11]\subset \mathbb{N}$, and the weights of uniform, Manhattan distance, or Euclidean distance. 
The best model had 6 neighbours in Euclidean distance.

\paragraph{GNB} The model did not have a tunable hyper-parameter.

\paragraph{SVM} The model was tuned on the kernel type in $\{\text{linear, poly, rbf, sigmoid}\}$, regularization parameter $C$ in range $[0.1, 2.0]\subset \mathbb{R}$, kernel coefficient gamma in $\{\text{scale, auto}\}$, and degree of the polynomial kernel function in range $[2,4]\subset \mathbb{N}$.
The best model used RBF kernel with scaled weights and regularization parameter of 1.8.

\paragraph{RF} The model did not restrict on the maximum depth of the trees. It was tuned on the class weight in settings of uniform, balanced, and balanced over sub-samples, and the minimum samples required to split a tree node in $\{2,7,12,...,97\}$.
The best model had balanced class weight and minimum of 17 samples to split a tree node.

\paragraph{Bagging} The model had 10 base estimators in the ensemble. It was tuned on the base estimator in SVM, Decision Tree, and KNN classifiers, and the proportion of maximum features used to train internal weak classifiers within range $[0.1, 1.0]\subset \mathbb{R}$.
The best model used maximum 50\% of all features to fit SVM as internal base estimator.

\paragraph{Voting} The model took the first six aforementioned trained models as inputs in the ensemble to vote for the output and was tuned on the choice of hard (voting on top-1 prediction) and soft (voting on the averaged logits) voting mechanism.
The best model used the soft voting mechanism.

\paragraph{Stacking} The model stacked the outputs of the first six aforementioned trained models in the ensemble followed by a final estimator and was tuned on the choice of final estimator among SVM and Logistic Regression.
The best model used Logistic Regression as the final estimator.

\section{Nomenclature}\label{sec:Nomenclature}
Table~\ref{T_nomenclature} and Table~\ref{T_func} give an overview of the mathematical notations used in this paper.

\section{Definition of Categories for Heritage Values and Attributes}\label{sec:Categories}
Table~\ref{T_OUV_definition} and Table~\ref{T_attr_definition} respectively give a detailed definition of heritage values (in terms of Outstanding Universal Value selection criteria) and heritage attributes (in terms of depicted scenes) categories applied in this paper.

\section{Multi-Graph Visualization}\label{sec:GraphVis}
The connected components of each type of temporal, social, and spatial links of each case study city are visualized in Figure~\ref{fig:graph_vis}, respectively.
The \texttt{spring\_layout} algorithm of NetworkX python library with optimal distance between nodes \texttt{k} of 0.1 and random seed of \texttt{10396953} are used to output the graphs.

\clearpage
\onecolumn

\renewcommand{\thetable}{A\arabic{table}}
\renewcommand{\thefigure}{A\arabic{figure}}

{\scriptsize\centering\sf
\captionsetup{width=\linewidth}
\begin{longtable}{rp{150px}p{260px}}
\caption{\footnotesize The nomenclature of mathematical notations used in this paper in alphabetic order. All superscripts of matrices are merely tags, not to be confused with exponents and operations, with the exception of transpose operator $\square^{\mathsf{T}}$.} \label{T_nomenclature}\\

\toprule \multicolumn{1}{c}{\textbf{Symbol}} & \multicolumn{1}{c}{\textbf{Data Type/Shape}} & \multicolumn{1}{c}{\textbf{Description}} \\ \midrule
\endfirsthead

\multicolumn{3}{c}%
{{\bfseries \tablename\ \thetable{} -- continued from previous page}} \\
\toprule
\multicolumn{1}{c}{\textbf{Symbol}} & \multicolumn{1}{c}{\textbf{Data Type/Shape}} & \multicolumn{1}{c}{\textbf{Description}} \\ \midrule 
\endhead

\midrule \multicolumn{3}{r}{{Continued on next page}} \\ \midrule
\endfoot

\bottomrule
\endlastfoot

$\boldsymbol{A}$ & Matrix of Boolean $\boldsymbol{A} := (\boldsymbol{A}^{\text{TEM}}>0)\bigvee (\boldsymbol{A}^{\text{SOC}}>0)\bigvee (\boldsymbol{A}^{\text{SPA}}>0) \in \{0,1\}^{K \times K}$ & The adjacency matrix of all post nodes in the set $\mathcal{V}$ that have at least one link connecting them as a composed simple graph.\\
$\boldsymbol{A}^{(*)}$ & Matrix of Float $\boldsymbol{A}^{(*)}:=[a^{(*)}_{i, i'}]_{K\times K} \in \mathbb{R}^{K \times K}$, $\boldsymbol{A}^{(*)}\in \{\boldsymbol{A}^\text{TEM},\boldsymbol{A}^\text{SOC},\boldsymbol{A}^\text{SPA}\}$ & The weighted adjacency matrix of each of the three sub-graphs $\mathcal{G}^{(*)}$ of the multi-graph $\mathcal{G}$, `(*)' represents one of the link types in \{TEM, SOC, SPA\}.\\
$\boldsymbol{A}^\mathcal{U}$ & Matrix of Boolean $\boldsymbol{A}^\mathcal{U}:=[a^\mathcal{U}_{j, j'}]_{|\mathcal{U}|\times |\mathcal{U}|} \in\{0,1\}^{|\mathcal{U}| \times |\mathcal{U}|}$ & The adjacency matrix of all unique users $\mathcal{U}$ marking their direct friendship which also included the relationship among themselves.\\
$\boldsymbol{A}^\mathcal{U'}$ & Matrix of Float $\boldsymbol{A}^\mathcal{U'}:=[a^\mathcal{U'}_{j, j'}]_{|\mathcal{U}|\times |\mathcal{U}|} \in[0,1]^{|\mathcal{U}| \times |\mathcal{U}|}$ & The weighted adjacency matrix of all unique users $\mathcal{U}$ marking their mutual interest in terms of the Jaccard Index of the public groups that they follow.\\
$\alpha_\mathcal{T}, \alpha_\mathcal{U}^{(n)}$ & Float scalars $\alpha_\mathcal{T}, \alpha_\mathcal{U}^{(1)}, \alpha_\mathcal{U}^{(2)}, \alpha_\mathcal{U}^{(3)} \in [0,1]$ & Parameters adjusting the weights of linear combination in relationship matrices $\boldsymbol{\mathfrak{T}}$ and $\boldsymbol{\mathfrak{U}}$.\\
$\beta_\mathcal{U}$ & Float scalar $\beta_\mathcal{U}\in (0,1)$ & The threshold to define mutual interest of two users as the Jaccard Index of public groups.\\
$\chi^2$ & Float Scalar & The Chi-square statistics of two distributions.\\
$\boldsymbol{\mathfrak{d}}_i,\mathfrak{D}$ & Object Tuples $\boldsymbol{\mathfrak{d}}_i = (\boldsymbol{\mathfrak{I}}_i, \mathcal{S}_i, \mathfrak{u}_i, \mathfrak{t}_i, \mathfrak{l}_i), \boldsymbol{\mathfrak{d}}_i \in \mathfrak{D}= \{\boldsymbol{\mathfrak{d}}_1, \boldsymbol{\mathfrak{d}}_2,..., \boldsymbol{\mathfrak{d}}_K\}$ & The tuple of all raw data (image, sentences, user ID, timestamp, and geo-location) from one sample point.\\
$D_\text{KL}$ & Float Scalar & The Kullback–Leibler (KL) divergence of two distributions.\\
$\boldsymbol{F}$ & Matrix of Integers and Floats $\boldsymbol{F}=[\boldsymbol{\mathfrak{f}}_i]_{3\times K}, \boldsymbol{\mathfrak{f}}_i = [\mathfrak{f}_{1,i},\mathfrak{f}_{2,i},\mathfrak{f}_{3,i}], \mathfrak{f}_{1,i}\in \mathbb{N}, \mathfrak{f}_{2,i},\mathfrak{f}_{3,i}\in [0,1]$ & The face recognition result of an image sample in terms of the number of faces detected $\mathfrak{f}_{1,i}$, the model confidence for the prediction $\mathfrak{f}_{2,i}$, and the proportion of total area of bounding boxes of detected faces to the total area of images $\mathfrak{f}_{3,i}$.\\
$G_0$ & Undirected weighted graph $G_0=(V_0,E_0,\boldsymbol{w}_0)$ & The complete spatial network in a city weighted by the travel time with all sorts of transportation between spatial nodes.\\
$G$ & Undirected weighted graph $G=(V,E,\boldsymbol{w}), V\subseteq V_0, E\subseteq E_0, \boldsymbol{w} \subseteq \boldsymbol{w}_0$ & The spatial network in a city weighted by the travel time between spatial nodes (no more than 20 minutes) that have at least one sample posted near them.\\
$\mathcal{G}$ & Weighted multi-graph $\mathcal{G}=(\mathcal{V}, \{\mathcal{E}^\text{TEM},\mathcal{E}^\text{SOC},\mathcal{E}^\text{SPA}\}, \{\boldsymbol{w}^\text{TEM},\boldsymbol{w}^\text{SOC},\boldsymbol{w}^\text{SPA}\})$ & The graph including the temporal, social, and spatial links $\mathcal{E}^*$ among the post nodes from set $\mathcal{V}$, weighted by the respective connection strengths $\boldsymbol{w}^*$.\\
$\mathcal{G}^{(*)}$ & Undirected weighted graph $\mathcal{G}^{(*)}=(\mathcal{V},\mathcal{E}^{(*)},\boldsymbol{w}^{(*)})$, $\mathcal{G}^{(*)}\in \{\mathcal{G}^\text{TEM},\mathcal{G}^\text{SOC},\mathcal{G}^\text{SPA}\}$ & The sub-graph of the multi-graph $\mathcal{G}$, while `(*)' represents one of the link types in \{TEM, SOC, SPA\}.\\
$\boldsymbol{H}^\text{B}$ & Matrix of Floats $\boldsymbol{H}^\text{B}=[\boldsymbol{h}^\text{BERT}_i]_{768\times K}$ & The last hidden layer for [CLS] token of BERT model pretrained on \emph{WHOSe\_Heritage}.\\
$\boldsymbol{H}^\text{v}$ & Matrix of Floats $\boldsymbol{H}^\text{v}=[\boldsymbol{h}^\text{v}_i]_{512\times K}$ & The last hidden layer of ResNet-18 model pretrained on \emph{Places365}.\\
$i,i'$ & Integer Indices $i,i'\in\{1,2,..,K\}\subset \mathbb{N}$ & The index of samples in the dataset $\mathfrak{D}$ of one case city.\\
$\boldsymbol{\mathfrak{I}}_i$ & Tensor of Integers within $[0,255]\in \mathbb{N}$ of size $150\times 150 \times 3$ or $320\times 240 \times 3$ & The raw image data of one sample post with RGB channels.\\
$\boldsymbol{I}$ & Matrix of Boolean $\boldsymbol{I}\in \{0,1\}^{|\mathcal{U}| \times |\mathcal{U}|}$ & The diagonal identity matrix marking the identity of unique users in $\mathcal{U}$.\\
$j,j'$ & Integer Indices $j,j'\in\{1,2,..,|\mathcal{U}|\}\subset \mathbb{N}$ & The index of users in the ordered set $\mathcal{U}$ of all unique users from one case city.\\
$k$ & Integer Indices $k\in\{1,2,..,|\mathcal{T}|\}\subset \mathbb{N}$ & The index of timestamps in the ordered set $\mathcal{T}$ of all unique timestamps from one case city.\\
$K$ & Integer $K=|\mathfrak{D}|$ & The sample size (number of posts) collected in one case city.\\
$\boldsymbol{K}^\text{HA}$ & Matrix of Floats $\boldsymbol{K}^\text{HA} = [\boldsymbol{\kappa}^\text{HA}_i]_{2\times K}$ & The confidence indicator matrix for heritage attributes labels including the top-$n$ confidence and agreement between VOTE and STACK models.\\
$\boldsymbol{K}^\text{HV}$ & Matrix of Floats $\boldsymbol{K}^\text{HV} = [\boldsymbol{\kappa}^\text{HV}_i]_{2\times K}$ & The confidence indicator matrix for heritage values labels including the top-$n$ confidence and agreement between BERT and ULMFiT models.\\
$l,l'$ & Integer Indices $l,l'\in\{1,2,..,|V|\}\subset \mathbb{N}$ & The index of nodes in the ordered set $V$ of all spatial nodes from one case city.\\
$\mathfrak{l}_i$ & Tuple of Floats $\mathfrak{l}_i = (\mathfrak{x}_i,\mathfrak{y}_i)$ & The geographical coordinate of latitude and longitude as location of one sample.\\
$\boldsymbol{L}^\text{a}$ & Matrix of logit vectors $\boldsymbol{L}^\text{a}=[\boldsymbol{l}^\text{a}_i]_{102\times K}$ & The last softmax layer of ResNet-18 model pretrained on \emph{SUN} predicting scene attributes.\\
$\boldsymbol{L}^\text{s}$ & Matrix of logit vectors $\boldsymbol{L}^\text{s}=[\boldsymbol{l}^\text{s}_i]_{365\times K}$ & The last softmax layer of ResNet-18 model pretrained on \emph{Places365} predicting scene categories.\\
$\mathcal{M}$ & A set of objects & The set of machine learning models used to train classifier on Tripoli data.\\
$\boldsymbol{O}$ & Matrix of Boolean $\boldsymbol{O}:=[\boldsymbol{\mathfrak{o}}_i]\in \{0,1\}^{3\times K}$ & The language detection result of the original language appearance of the sentences in each sample, in terms of English $\boldsymbol{\mathfrak{o}}_1$, local language $\boldsymbol{\mathfrak{o}}_2$, and other languages $\boldsymbol{\mathfrak{o}}_3$.\\
$\boldsymbol{R},\boldsymbol{R}^{(*)}$ & Matrix of Float $\boldsymbol{R},\boldsymbol{R}^{(*)}\in \mathbb{R}^{N \times K}$, $\boldsymbol{R}^{(*)}\in \{\boldsymbol{R}^\text{TEM},\boldsymbol{R}^\text{SOC},\boldsymbol{R}^\text{SPA}\}$ & The embedding matrices of each of the samples to a $N$-dimensional vector based on the general structure of the multi-graph $\mathcal{G}$ and the specific types of links.\\
$\mathcal{S}_i$ & Set of Strings $\mathcal{S}_i=\{\mathcal{s}^{(1)}_i, \mathcal{s}^{(2)}_i, ..., \mathcal{s}^{(|\mathcal{S}_i|)}_i\}$ or Empty Set $\mathcal{S}_i=\varnothing$ & The processed textual data as a set of individual sentences that have a valid semantic meaning and have been translated into English.\\
$\boldsymbol{S}$ & Boolean Matrix $\boldsymbol{S}:=[s_{l,i}]\in \{0,1\}^{|V| \times K}$ & The one-hot embedding matrix of the samples corresponding to the geo-node set $V$.\\
$\boldsymbol{\mathfrak{S}}$ & Matrix of Float $\boldsymbol{\mathfrak{S}}:=[\mathfrak{s}_{l,l'}]\in [0,1]^{|V|\times |V|}$ & A matrix marking the spatial closeness of all the unique spatial nodes from set $V$ that can be reached within 20 minutes.\\
$\mathcal{T}$ & An ordered Set $\mathcal{T}=\{\tau_1, \tau_2, ..., \tau_{|\mathcal{T}|}\}$ & The ordered set of all unique timestamps from one case city.\\
$\tau_k$ & Timestamp $\tau_k\in \mathcal{T}$ & A timestamp in the ordered set $\mathcal{T}$ of all unique timestamps.\\
$\mathfrak{t}_i$ & Timestamp $\mathfrak{t}_i\in \mathcal{T}$ & A timestamp indexed with sample ID in the ordered set $\mathcal{T}$ of all unique timestamps.\\
$\boldsymbol{T}$ & Boolean Matrix $\boldsymbol{T}:=[t_{k,i}]\in \{0,1\}^{|\mathcal{T}| \times K}$ & The one-hot embedding matrix of the samples corresponding to the timestamp set $\mathcal{T}$.\\
$\boldsymbol{\mathfrak{T}}$ & Matrix of Float $\boldsymbol{\mathfrak{T}}\in [0,1]^{|\mathcal{T}|\times |\mathcal{T}|}$ & A matrix marking the temporal similarity of all the unique timestamps from set $\mathcal{T}$.\\
$\mathcal{U}$ & An ordered Set $\mathcal{U}=\{\mu_1, \mu_2, ..., \mu_{|\mathcal{U}|}\}$ & The ordered set of all unique users from one case study city.\\
$\mu_j$ & User ID Object $\mu_j\in \mathcal{U}$ & An instance of user in the ordered set $\mathcal{U}$ of all unique users.\\
$\mathfrak{u}_i$ & User ID Object $\mathfrak{u}_i\in \mathcal{U}$ & An instance of user indexed with sample ID in the ordered set $\mathcal{U}$ of all unique users.\\
$\boldsymbol{U}$ & Boolean Matrix $\boldsymbol{U}:=[u_{j,i}]\in \{0,1\}^{|\mathcal{U}| \times K}$ & The one-hot embedding matrix of the samples corresponding to the user set $\mathcal{U}$.\\
$\boldsymbol{\mathfrak{U}}$ & Matrix of Float $\boldsymbol{\mathfrak{U}}\in [0,1]^{|\mathcal{U}|\times |\mathcal{U}|}$ & A matrix marking the social similarity of all the unique users from set $\mathcal{U}$, as a linear combination of identity matrix $\boldsymbol{I}$ and adjacency matrices $\boldsymbol{A}^\mathcal{U}, \boldsymbol{A}^\mathcal{U'}$.\\
$V$ & A set of nodes $V=\{\upsilon_1, \upsilon_2, ..., \upsilon_{|V|}\}$ & The set of all the spatial nodes that have at least one sample posted near them.\\
$\upsilon_l$ & Spatial node $\upsilon_l\in V$ & A node in the set $V$ of all spatial nodes that have at least one sample posted near them.\\
$\mathcal{V}$ & A set of nodes $\mathcal{V}=\{v_1, v_2, ..., v_K\}$ & The set of all nodes of posts in one case city.\\
$v_i$ & Post/Sample node $v_i\in \mathcal{V}$ & A node in the set $\mathcal{V}$ of all nodes of posts in one case city.\\
$\boldsymbol{w},\boldsymbol{w}^{(*)}$ & Vector of Float $\boldsymbol{w}:=[w_e]\in [0,20]^{|E|}, \boldsymbol{w}^{(*)}:=[w^{(*)}_e]\in \mathbb{R}^{|\mathcal{E}|}, \boldsymbol{w}^{(*)}\in\{\boldsymbol{w}^\text{TEM}, \boldsymbol{w}^\text{SOC}, \boldsymbol{w}^\text{SPA}\}$ & The weight vector of spatial network $G$ and post graphs $\mathcal{G}^\text{TEM},\mathcal{G}^\text{SOC},\mathcal{G}^\text{SPA}$, these weights are directly interchangeable with the adjacency matrices.\\
$\boldsymbol{X}^\text{vis}$ & Matrix of Floats and Integers $\boldsymbol{X}^\text{vis}_{982\times K} = {\left[{\boldsymbol{H}^\text{v}}^\mathsf{T}, {\boldsymbol{F}}^\mathsf{T}, {\boldsymbol{\sigma}^{(5)}(\boldsymbol{L}^\text{s})}^\mathsf{T}, {\boldsymbol{\sigma}^{(10)}(\boldsymbol{L}^\text{a})}^\mathsf{T}\right]}^\mathsf{T}$ & The final visual feature concatenating the hidden layer $\boldsymbol{H}^\text{v}$, the face detection results $\boldsymbol{F}$, the filtered top-5 scene prediction $\boldsymbol{\sigma}^{(5)}(\boldsymbol{L}^\text{s})$, and the filtered top-10 attribute prediction $\boldsymbol{\sigma}^{(10)}(\boldsymbol{L}^\text{a})$.\\
$\boldsymbol{X}^\text{tex}$ & Matrix of Floats and Integers $\boldsymbol{X}^\text{tex}_{771\times K} = {\left[{\boldsymbol{H}^\text{B}}^\mathsf{T}, {\boldsymbol{O}}^\mathsf{T}\right]}^\mathsf{T}$ & The final textual feature concatenating the hidden layer $\boldsymbol{H}^\text{B}$ of BERT on [CLS] token, and the original language detection results $\boldsymbol{O}$.\\
$\boldsymbol{Y}^\text{HA}$ & Matrix of Floats $\boldsymbol{Y}^\text{HA} = [\boldsymbol{y}^\text{HA}_i]_{9\times K}$ & The final generated label of heritage attributes on 9 depicted scenes, as the average of prediction from VOTE and STACK models.\\
$\boldsymbol{Y}^\text{HV}$ & Matrix of Floats $\boldsymbol{Y}^\text{HV} = [\boldsymbol{y}^\text{HV}_i]_{11\times K}$ & The final generated label of heritage values on 10 OUV selection criteria and an additional negative class, as the average of prediction from BERT and ULMFiT models.\\
\end{longtable}
}

\begin{table*}[ht]
\scriptsize\centering
\caption{\label{T_func}
\footnotesize The nomenclature of functions defined and used in this paper in alphabetic order.}
\sf\begin{tabular}{rp{125px}p{200px}}
\toprule
Symbol & Data Type/Shape & Description\\ 
\midrule
$\text{argmx}(\boldsymbol{l},n)$ & Function outputting a Set of floats or objects & The set of largest $n$ elements of any float vector $\boldsymbol{l}$.\\
$\boldsymbol{f}_\text{BERT}(\mathcal{S}|\boldsymbol{\Theta}_\text{BERT})$ & Function inputting a sentence/paragraph or a batch of sentences/paragraphs, outputting a vector or a matrix of vectors & The pretrained uncased BERT model fine-tuned on \emph{WHOSe\_Heritage} with the model parameters $\boldsymbol{\Theta}_\text{BERT}$ that can process some textual inputs into the 768-dimensional hidden output vector $\boldsymbol{h}^\text{BERT}$ of the \texttt{\tiny [CLS]} token.\\
$\boldsymbol{f}_\text{ResNet-18}(\boldsymbol{\mathfrak{I}}|\boldsymbol{\Theta}_\text{ResNet-18})$ & Function inputting a tensor or a batch of tensors, outputting three vectors or three matrices of vectors & The ResNet-18 model pretrained on \emph{Places365} dataset with the model parameters $\boldsymbol{\Theta}_\text{ResNet-18}$ that can process the image tensor $\boldsymbol{\mathfrak{I}}$ into the predicted vectors of scenes $\boldsymbol{l}^\text{s}$, predicted vectors of attributes $\boldsymbol{l}^\text{a}$, and the last hidden layer $\boldsymbol{h}^\text{v}$.\\
$\boldsymbol{g}_\text{BERT}(\mathcal{S}|\boldsymbol{\Theta}_\text{BERT})$& Function inputting a sentence/paragraph or a batch of sentences/paragraphs, outputting a vector or a matrix of vectors & The end-to-end pretrained uncased BERT model fine-tuned on \emph{WHOSe\_Heritage} with the model parameters $\boldsymbol{\Theta}_\text{BERT}$ together with the MLP classifiers that can process some textual inputs into the logit prediction vector $\boldsymbol{y}^\text{BERT}$ of 11 heritage value classes concerning OUV.\\
$\boldsymbol{g}_\text{ULMFiT}(\mathcal{S}|\boldsymbol{\Theta}_\text{ULMFiT})$& Function inputting a sentence/paragraph or a batch of sentences/paragraphs, outputting a vector or a matrix of vectors & The end-to-end pretrained ULMFiT model fine-tuned on \emph{WHOSe\_Heritage} with the model parameters $\boldsymbol{\Theta}_\text{ULMFiT}$ together with the MLP classifiers that can process some textual inputs into the logit prediction vector $\boldsymbol{y}^\text{ULMFiT}$ of 11 heritage value classes concerning OUV.\\
$\boldsymbol{h}_\text{VOTE}(\boldsymbol{h}^\text{v}|{\Theta}_\text{VOTE},\mathcal{M},\boldsymbol{\Theta}_\mathcal{M})$ & Function inputting a vector or a batch of vectors, outputting a vector or a matrix of vectors & The ensemble Voting Classifier with model parameter ${\Theta}_\text{VOTE}$ of machine learning models from $\mathcal{M}$ with their respective model parameters $\boldsymbol{\Theta}_\mathcal{M}$, which processes the visual feature vector $\boldsymbol{h}^\text{v}$ into the logit prediction vector $\boldsymbol{y}^\text{VOTE}$ of 9 heritage attribute classes concerning depicted scenes.\\
$\boldsymbol{h}_\text{STACK}(\boldsymbol{h}^\text{v}|{\Theta}_\text{STACK},\mathcal{M},\boldsymbol{\Theta}_\mathcal{M})$ & Function inputting a vector or a batch of vectors, outputting a vector or a matrix of vectors & The ensemble Stacking Classifier with model parameter ${\Theta}_\text{STACK}$ of machine learning models from $\mathcal{M}$ with their respective model parameters $\boldsymbol{\Theta}_\mathcal{M}$, which processes the visual feature vector $\boldsymbol{h}^\text{v}$ into the logit prediction vector $\boldsymbol{y}^\text{STACK}$ of 9 heritage attribute classes concerning depicted scenes.\\
$\mathcal{I}(\mu_j)$ & Function outputting an ordered Set of Objects & The set of public groups that are followed by user $\mu_j$.\\
$\text{IoU}(\mathcal{A},\mathcal{B})$ & Function outputting a positive float & The Jaccard Index of any two sets $\mathcal{A},\mathcal{B}$ as the cardinality of the intersection of the two sets over that of the union of them.\\
$\text{max}(\boldsymbol{l},n)$ & Function outputting a Float & The $n_\text{th}$ largest element of any float vector $\boldsymbol{l}$.\\
$\boldsymbol{\sigma}^{(n)}(\boldsymbol{l})$ & Function both inputting and outputting a Logit Vector & The activation filter to keep the top-$n$ entries of any logit vector $\boldsymbol{l}$ and smooth all the others entries based on the total confidence of top-$n$ entries.\\
\bottomrule
\end{tabular}
\end{table*}

\begin{table*}[ht]
\scriptsize\centering
\caption{\label{T_OUV_definition}
\footnotesize The definition for each UNESCO World Heritage OUV selection criterion as \textbf{heritage value category} in this dataset and its main topic according to \cite{UNESCO2008}, \cite{Jokilehto2008}, and \cite{Bai2021}.}
\sf\begin{tabular}{clp{360pt}}
\toprule
Criterion & Focus & Definition \\ 
\midrule
(i) & Masterpiece & \emph{To represent a masterpiece of human creative genius};\\
(ii) & Values/Influence &\emph{To exhibit an important interchange of human values, over a span of time or within a cultural area of the world, on developments in architecture or technology, monumental arts, town-planning or landscape design};\\
(iii) & Testimony &\emph{To bear a unique or at least exceptional testimony to a cultural tradition or to a civilization which is living or which has disappeared};\\
(iv) & Typology &\emph{To be an outstanding example of a type of building, architectural or technological ensemble or landscape which illustrates (a) significant stage(s) in human history};\\
(v) & Land-Use &\emph{To be an outstanding example of a traditional human settlement, land-use, or sea-use which is representative of a culture (or cultures), or human interaction with the environment especially when it has become vulnerable under the impact of irreversible change};\\
(vi) & Associations &\emph{To be directly or tangibly associated with events or living traditions, with ideas, or with beliefs, with artistic and literary works of outstanding universal significance};\\
(vii) & Natural Beauty&\emph{To contain superlative natural phenomena or areas of exceptional natural beauty and aesthetic importance};\\
(viii) & Geological Process&\emph{To be outstanding examples representing major stages of earth's history, including the record of life, significant on-going geological processes in the development of landforms, or significant geomorphic or physiographic features};\\
(ix) & Ecological Process&\emph{To be outstanding examples representing significant on-going ecological and biological processes in the evolution and development of terrestrial, fresh water, coastal and marine ecosystems and communities of plants and animals};\\
(x) & Bio-diversity &\emph{To contain the most important and significant natural habitats for in-situ conservation of biological diversity, including those containing threatened species of outstanding universal value from the point of view of science or conservation}.\\
\bottomrule
\end{tabular}
\end{table*}

\begin{table*}[ht]
\scriptsize\centering
\caption{\label{T_attr_definition}
\footnotesize The definition for depicted scenery as \textbf{heritage attribute category} in this dataset and its tangible/intangible type according to \cite{Veldpaus2015}, \cite{gustcoven2016attributes}, and \cite{Ginzarly2019}.}
\sf\begin{tabular}{p{90pt}lp{290pt}}
\toprule
Attribute & Type & Definition \\ 
\midrule
Monuments and Buildings & Tangible &\emph{The exterior of a whole building, structure, construction, edifice, or remains that host(ed) human activities, storage, shelter or other purpose};\\
Building Elements & Tangible & \emph{Specific elements, details, or parts of a building, which can be constructive, constitutive, or decorative};\\
Urban Form Elements & Tangible &\emph{Elements, parts, components, or aspects of/in the urban landscape, which can be a construction, structure, or space, being constructive, constitutive, or decorative};\\
Urban Scenery & Tangible &\emph{A district, a group of buildings, or specific urban ensemble or configuration in a wider (urban) landscape or a specific combination of cultural and/or natural elements};\\
Natural Features and Landscape Scenery & Tangible &\emph{Specific flora and/or fauna, such as water elements of/in the historic urban landscape produced by nature, which can be natural and/or designed};\\
Interior Scenery & Tangible/ Intangible &\emph{The interior space, structure, construction, or decoration that host(ed) human activity, showing a specific (typical, common, special) use or function of an interior place or environment};\\
People's Activity and Association & Intangible &\emph{Human associations with a place, element, location, or environment, which can be shown with the activities therein};\\
Gastronomy & Intangible &\emph{The (local) food-related practices, traditions, knowledge, or customs of a community or group, which may be associated with a community or society and/or their cultural identity or diversity};\\
Artifact Products & Intangible &\emph{The (local) artifact-related practices, traditions, knowledge, or customs of a community or group, which may be associated with a community or society and/or their cultural identity or diversity}.\\
\bottomrule
\end{tabular}
\end{table*}

\begin{figure*}[ht]
\centering
\includegraphics[width=0.85\linewidth]{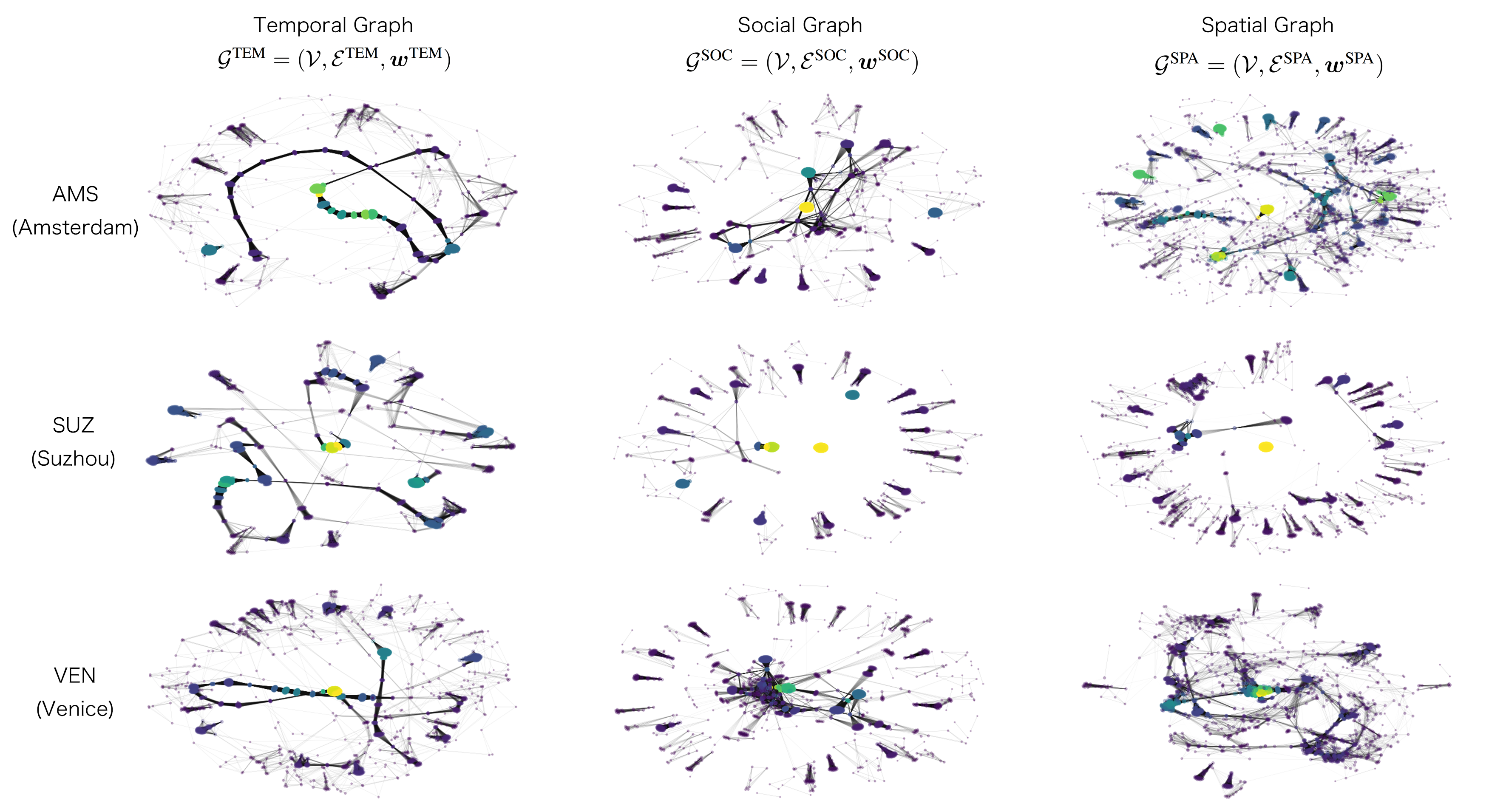}
\caption{\footnotesize The subgraphs of the multi-graphs in each case study city visualized using spring layout in NetworkX. The node size and colour reflect the degrees, and link thickness the edge weights.}\label{fig:graph_vis}
\end{figure*}

\end{document}